\documentclass{jfm}

\usepackage{graphicx}
\usepackage{natbib}

\ifCUPmtlplainloaded \else
  \checkfont{eurm10}
  \iffontfound
    \IfFileExists{upmath.sty}
      {\typeout{^^JFound AMS Euler Roman fonts on the system,
                   using the 'upmath' package.^^J}%
       \usepackage{upmath}}
      {\typeout{^^JFound AMS Euler Roman fonts on the system, but you
                   dont seem to have the}%
       \typeout{'upmath' package installed. JFM.cls can take advantage
                 of these fonts,^^Jif you use 'upmath' package.^^J}%
      }
  \else
  \fi
\fi

\ifCUPmtlplainloaded \else
  \checkfont{msam10}
  \iffontfound
    \IfFileExists{amssymb.sty}
      {\typeout{^^JFound AMS Symbol fonts on the system, using the
                'amssymb' package.^^J}%
       \usepackage{amssymb}%
         \let\leq=\leqslant
         \let\geq=\geqslant
      }{}
  \fi
\fi

\ifCUPmtlplainloaded \else
  \IfFileExists{amsbsy.sty}
    {\typeout{^^JFound the 'amsbsy' package on the system, using it.^^J}%
     \usepackage{amsbsy}}
    {}
\fi

\newsavebox{\astrutbox}
\sbox{\astrutbox}{\rule[-5pt]{0pt}{20pt}}

\newcommand\p{\ensuremath{\partial}}

\usepackage{graphicx,epsfig,epstopdf,dsfont,color}\usepackage{soul}
\usepackage{amsmath,amssymb,graphicx}
\usepackage{bm,gensymb}
\usepackage[dvipsnames]{xcolor}

\def\red{\textcolor{red}}
\def\blue{\textcolor{blue}}

\def\yellow{\textcolor{BurntOrange}}
\definecolor{darkblue}{RGB}{83,0,93}
\def\L{\mbox{------}}

\def\I{\mathcal{I}}
\def\d{\text{d}}
\def\bes{\begin{eqnarray}}
\def\ees{\end{eqnarray}}
\def\beq{\begin{equation}}
\def\eeq{\end{equation}}
\def\ba#1\ea{\begin{align}#1\end{align}}
\def\bsa#1#2\esa{\begin{subequations}\label{#1}
\begin{align}#2\end{align} \end{subequations}}
\def\lp{\left(}
\def\rp{\right)}
\def\lb{\left[}
\def\rb{\right]}
\def\lcb{\left\{}
\def\rcb{\right\}}

\def\f{\frac}
\def\b{\mathbf}
\def\p{\partial}
\def\co{{\mathcal O}}
\def\d{\textrm d}

\def\x{\bar{x}}
\def\y{\bar{y}}

\def\rm{r_{\pm m}}

\def\Am{A_{\pm m}}

\def\tm{\theta_{\pm m}}

\def\An{A_{\pm n}}
\def\tn{\theta_{\pm n}}

\def\ka{\kappa}
\def\kain{\kappa_{in}}
\def\Kain{\bm{\kappa}_{in}}
\def\kab{\kappa_b}
\def\Kab{\bm{\kappa}_b}
\def\kaj{\kappa_j}
\def\Kaj{\bm{\kappa}_j}
\def\kam{\kappa_{r_{\pm m}}}
\def\Kam{\bm{\kappa}_{r_{\pm m}}}

\def\Kan{\bm{\kappa}_{r_{\pm n}}}

\def\dm{\delta_{\pm m}}
\def\Dm{\bm{\delta}_{\pm m}}
\def\D2m{\bm{\Delta}_{\pm m}}

\def\uh{\textbf{u}_h}
\def\xh{\textbf{x}_h}
\def\sxh{\overline{\textbf{x}}_h}
\def\snab{\overline{\nabla}_h}
\def\kb{\textbf{k}_b}

\def\blue{\textcolor{blue}}

\title[Oblique internal-wave chain resonance over seabed corrugations]{Oblique internal-wave chain resonance over seabed corrugations}

\author[L.-A. Couston, Y.  Liang and M.-R. Alam]{Louis-Alexandre Couston$^{1,2}$, Yong Liang$^3$ and Mohammad-Reza Alam$^{1,3}$\thanks{reza.alam@berkeley.edu}
}

\affiliation{$^1$ Department of Mechanical Engineering, University of California, Berkeley, CA 94720\\ $^2$ CNRS, Aix Marseille Univ, Centrale Marseille, IRPHE, Marseille, France\\ $^3$ Applied Science and Technology, University of California, Berkeley, CA 94720, USA}

\usepackage{amsmath}

\begin{document}
\date{\today}
\maketitle

%%%%%%%%%%%%%%%%%%%%%%
\begin{abstract}
Here we show that monochromatic long-crested corrugations on an otherwise flat seafloor can coherently scatter the energy of an oblique incident  internal wave to multiple multi-directional higher-mode internal waves via a series of resonant interactions. 
We demonstrate that a resonance between seabed corrugations and a normally or slightly oblique incident internal wave results in a series of follow-up resonant interactions, which take place between the same corrugations and successively resonated shorter waves. A chain resonance of internal waves that carries energy to small scales is thus obtained, and we find that the Richardson number decreases by several orders of magnitude over the corrugated patch.
If the incidence angle is large, and the incident wave perfectly satisfies a resonance condition with the topography, it turns out that not many higher-mode resonance or near-resonance conditions can be satisfied, such that energy stays confined within the first few modes. 
Nevertheless, if the incident waves are sufficiently detuned from satisfying a perfect resonance condition with the seabed corrugations, then we show that this frequency detuning may balance off the large detuning due to oblique incidence, leading to a chain resonance that again carries energy to small scales. 
The evolution of the incident and resonated wave amplitudes is predicted from the envelope equation for  internal waves over resonant seabed topography in a three-dimensional rotating fluid, which we derive considering the Boussinesq and $f$-plane approximations with $f$ the Coriolis frequency, linear density stratification and small-amplitude corrugations. 
Our results suggest that topographic features on the ocean floor with a well-defined dominant wavenumber vector, through the chain resonance mechanism elucidated here, may play a more important role than previously thought in the enhancement of diapycnal mixing and energy dissipation. 

\end{abstract}
%%%%%%%%%%%%%%%%%%%%%%

%%%%%%%%%%%%%%%%%%%%%%
\section{Introduction}
%%%%%%%%%%%%%%%%%%%%%%

The global oceanic circulation is believed to be maintained by the occurrence of enhanced diapycnal mixing events over some regions of the oceans \cite[][]{Munk1998}. The enhancement is expected to result from strong internal-wave activity, which, in the open ocean, has been found by field observations  to be most significant over underwater topographical features \cite[e.g. continental shelves, mid-oceanic ridges; cf.][]{Polzin1997,Ledwell2000,Nash2007,Waterhouse2014}. %Field observations have shown that high levels of mixing are often found over topographical features of the seafloor (e.g. continental shelves, mid-oceanic ridges), and pointed strong internal-wave activity as the primary source of mixing \cite[e.g.][]{Polzin1997,Ledwell2000,Nash2007,Waterhouse2014}. 
Energy injection into the internal-wave field \cite[][]{Wunsch2004} and the contribution of internal waves to turbulent mixing through nonlinear processes \cite[][]{Staquet2002} are thus important aspects of the ocean dynamics that are being gradually incorporated in global ocean models \cite[][]{Exarchou2012,Lefauve2015,Timko2017}.

The final processes by which internal waves lead to turbulent diapycnal mixing are fundamentally nonlinear \cite[][]{Staquet2002}. However, a significant fraction of energy injection in the internal-wave field first occurs via bottom scattering \cite[][]{Garrett2007}, which can be considered as a linear process when the bottom slope is subcritical \cite[see e.g.][for a case with critical slope]{Legg2014}. As a result, the linear generation of internal waves over variable seabed is of significant interest, and, for small-amplitude topography, some of the mechanisms that can be and have been studied using regular perturbation methods  include the topographic scattering of existing low-wavenumber internal waves toward higher wavenumbers \cite[][]{Muller1992}, the generation of topographic lee waves \cite[][]{Bell1975b}, and the conversion of barotropic tidal energy into internal waves over rough topography \cite[][]{Balmforth2002,Garrett2007}. The generation of internal waves by large-amplitude subcritical topography has, on the other hand, been investigated using the semi-analytical Green function approach \cite[][]{Buhler2007,Mathur2014} and ray tracing \cite[][]{Buhler2011,Guo2016}.

The generation of high-wavenumber internal waves by topographic scattering is of significant interest because nonlinear processes leading to turbulent mixing are more efficient when internal waves have short wavelengths \cite[][]{Sarkar2017}. Thus, a natural question is: what topographical features of the seafloor can most efficiently generate high-wavenumber waves? In the case of small-amplitude topography, previous studies have usually considered a continuous topographic spectrum and have shown that the amplitude of the generated waves is proportional to the height of topographical features \cite[][]{Bell1975,Muller1992,StLaurent2002,Llewellyn2002,Khatiwala2003}. This means that the bottom forcing by small seabed features is in general expected to result in a weak perturbation of the incident wavefield, such that only a small fraction of the available energy can be transferred to high-wavenumber internal waves. 

Because previous studies have focused on the overall effect of a continuous seabed spectrum, the possibility that some seabed wavenumbers (out of the complete spectrum) can resonantly generate  high-wavenumber internal waves has been seldom discussed (here we use the term resonant to qualify the generation of internal waves whose amplitudes are comparable to the amplitude of the incident wave). In a relatively recent investigation of wave focusing in a one-horizontal dimension finite-depth ocean, and in agreement with earlier analyses of tidal conversion \cite[][]{Garrett2007}, \cite{Buhler2011} yet briefly discussed that the interaction of an incident internal tide (wavenumber $k_1$) with a single bottom harmonic (wavenumber $k_b$) can  generate a series of shorter internal waves resonantly provided that $k_b/k_1$ is an integer. When $k_b=k_1$, all internal-wave wavenumbers are indeed integer multiples of $k_1$ such that each successively resonated wave $k_j$, starting with $k_2=k_1+k_b=2k_1$, can resonate other new higher-wavenumber internal waves $k_{j+1}=k_j+k_b$. Thus, the incident-wave energy is transferred to the shorter waves through perfect wave-wave resonances occurring in series, but the mechanism requires a two-dimensional linearly-stratified fluid and a top rigid lid.

In this paper, we demonstrate that the strong resonant transfer of energy from a low-mode internal wave to shorter waves by bottom scattering is not limited to the special two-dimensional case discussed in \cite{Buhler2011}, and that it can occur in three dimensions with oblique incidence of the low-mode wave at and away from perfect resonance  (cf. figure \ref{resodiag}). By focusing our analysis on monochromatic corrugations rather than bathymetric spectra, we can explore in detail the resonance mechanism and the energy transfers from the incident low-mode internal waves to the shorter waves. The monochromatic corrugations considered in this paper may be seen as idealizations of ridges and valleys with length scales 10-100 km running perpendicular to the prominent mid-ocean ridges, since a dominant bottom wavenumber vector can be extracted for these topographical features \cite[see e.g.][in which significant mixing is observed above crest-parallel bathymetric sills on a mid-ocean rigde flank]{Thurnherr2005}\footnote{See also available seafloor maps at http://topex.ucsd.edu/marine\_topo.}. Monochromatic corrugations cannot be considered representative of abyssal hills, however, because the topographic spectra of abyssal hills are assumed to have random phase distributions, and hence cannot be accurately approximated by a single bottom mode  \cite[we refer the reader to][for additional information on seafloor topography]{Goff1988,Smith1997,Becker2009,Goff2010}.

We solve the resonant three-dimensional boundary-value problem of internal-wave scattering at the steady-state by deriving an evolution equation for the spatial variations of envelope amplitudes using the method of multiple scales in the weak-topography limit. The method has been successfully applied for many investigations of wave-bottom resonance in homogeneous water (known as Bragg resonance), including shore protection \cite[][]{Yu2000a,Couston2017}, sandbar formation \cite[][]{Yu2000b}, wave lensing \cite[][]{Elandt2014}, and in two-layer density stratified fluids \cite[][]{Alam2009} with applications in the attenuation of long interfacial waves by random bathymetry \cite[][]{Alam2007a} and broadband cloaking \cite[][]{Alam2012}. Multiple-scale methods have also been employed in studies of wave dynamics in stratified fluids, including weakly nonlinear wave-mean flow interactions at topography \cite[][]{Nikurashin2010a} and the scattering of internal tides by irregular bathymetry \cite[][]{Li2014}. Because of the obliquity of the incident waves, but also because we allow the first interaction between the incident waves and the corrugations to be detuned, it should be noted that the chain resonances include mostly near-resonance waves. Near-resonance interactions (also known as detuned resonance) are of increasing interest in nonlinear wave science \cite[][]{Tobisch2016}, and will be shown to play a major role in the present analysis.

The outline of the paper is as follows. In \S\ref{sec:bi2}, we present the governing equations, characterize the different waves involved in chain resonance for arbitrary incidence angle and define a detuning parameter for each of the near-resonance interactions. In \S\ref{sec:bi3}, we provide the derivation details for the multiple-scale equation and give the solution method for the boundary-value problem. In \S\ref{sec:bi4}, we consider physical parameters representative of idealized oceanic conditions, and we describe the chain resonance results obtained for a wide range of incidence angles and detunings. Finally, in \S\ref{sec:bi5}, we provide concluding remarks.

%%%%%%%%%%%%%%%%%%%%%%
\section{Problem formulation}\label{sec:bi2}
%%%%%%%%%%%%%%%%%%%%%%

%%%%%%%%%%%%%%%%%%%%%%
\subsection{Governing equations}\label{sec:bi21}
%%%%%%%%%%%%%%%%%%%%%% 
 
Consider the propagation of small-amplitude waves in a stably stratified rotating fluid with constant buoyancy frequency $N$ and under $f$-plane approximation with $f$ the Coriolis frequency. In the Boussinesq approximation and neglecting viscosity, the linearized wave equation and boundary conditions for the vertical velocity $w$ in Cartesian domain $(x,y,z)$ can be written as \cite[e.g.][]{Sutherland2010}
\bsa{bi2}\label{bi2a}
 \lb (\p_t^2+f^2)\p_z^2+(\p_t^2+N^2)\nabla_h^2 \rb  w &= 0,~ &-H+b \leq z \leq 0, \\\label{bi2b}
 w &=0,~&z=0, \\\label{bi2c}
 w &=\uh\cdot\nabla_hb,~&z=-H+b,
\esa
where $\uh=(u,v)$ is the horizontal velocity vector, the $z$ axis is taken vertically upward with $z=0$ the top impermeable fixed boundary obtained under rigid-lid assumption, $b(x,y)$ represents the bottom variations from the mean water depth $H$ and $\nabla_h=(\p_x,\p_y)$ is the horizontal gradient. The horizontal velocities $(u,v)$ in equation \eqref{bi2c} can be expressed in terms of $w$ as 
\ba\label{pol}
\p_t\nabla_h^2 u = -(\p_t\p_x+f\p_y)\p_z w, ~
\p_t\nabla_h^2 v = -(\p_t\p_y-f\p_x)\p_z w,
\ea
following a straightforward manipulation of the continuity and linearized momentum equations, with equations \eqref{pol} known as the polarization relations for linear internal waves \cite[cf. e.g.][]{Sutherland2010}.
%In equations \eqref{bi2}-\eqref{pol}, the $z$ axis is taken vertical upward with $z=0$ the top impermeable fixed boundary obtained under rigid-lid assumption, $b(x,y)$ represents the bottom variations from the mean water depth $H$ and $\nabla_h=(\p_x,\p_y)$ is the horizontal gradient.  

Our goal is to investigate the effect of a finite patch of monochromatic seabed corrugations on incident internal waves coming from $x=-\infty$ at the steady state. Thus we consider a seabed topography superimposed on the mean water depth of the form
\beq\label{bi3}
b(x,y)=\lcb\begin{array}{c}
\f{d}{2} \lp e^{i\varphi}e^{i\kb\cdot \xh} + c.c. \rp,~0\leq x \leq \ell, -\infty \leq y \leq +\infty, \\
0,~\text{otherwise},
\end{array}\right.
\eeq 
with $\kb$ the corrugation wavenumber vector, $\varphi$ an arbitrary phase, $d$ the corrugation amplitude, $\ell$ the length of the patch, $c.c.$ denotes the complex conjugate and $\xh=(x,y)$. We look for a steady-state time-periodic solution of the form $(u,v,w)=Re[(\hat{u},\hat{v},\hat{w})e^{-i\omega t}]$ ($Re$ denoting the real part) with $\omega$ the frequency of the incident waves. Therefore, $\p_t$ can be substituted with $-i\omega$ and $(u,v,w)$ with $Re[(\hat{u},\hat{v},\hat{w})e^{-i\omega t}]$ in \eqref{bi2}-\eqref{pol}, leading to (dropping the hat symbol)
\bsa{dless}\label{dlessa}
 \lp \mu^2\p_z^2-\nabla_h^2 \rp  w &= 0,~ &-H+b \leq z \leq 0, \\\label{dlessb}
 w &=0,~&z=0, \\\label{dlessc}
 w &=\uh\cdot\nabla_hb,~&z=-H+b,
\esa
and
\ba\label{poless}
\nabla_h^2 u = -\lp\p_x+\f{if}{\omega}\p_y\rp\p_z w, ~
\nabla_h^2 v = -\lp\p_y-\f{if}{\omega}\p_x\rp\p_z w,
\ea
with $\b{u}(x,y,z)=(u,v,w)$ the time-independent complex-valued velocity field to be found and $\mu(\omega)=\sqrt{(\omega^2-f^2)/(N^2-\omega^2)}$ the so-called internal-waves slope \cite[e.g.][]{Buhler2011}.

Upstream ($x\leq 0$) and downstream ($x\geq \ell$) of the corrugated patch, the bottom is flat ($b=0$) and the solution of equation \eqref{bi2} can be written as a sum $w=\sum_j w_j$ of internal-wave modes of the form
\ba{}\label{sc0}
w_j=A_j\sin \lb \gamma_j (z+H)  \rb e^{i\b{k}_j\cdot\xh},
\ea
provided that $f<\omega<N$, with $\gamma_j=j\pi/H$ the vertical wavenumber and $A_j$ the complex wave-mode amplitude. For each wave mode \eqref{sc0}, the magnitude of the horizontal wavenumber $k_j=|\b{k}_j|$ can be related to the vertical wavenumber through the dispersion relation
\ba{}\label{bi8}
k_j=\mu\gamma_j=\mu \f{j\pi}{H}.
\ea

While the wave modes \eqref{sc0} propagate independently from each other over the flat bottom, they become coupled over the corrugations (i.e. when $b\neq 0$) through the bottom boundary condition \eqref{dlessc}. We will demonstrate that energy transfers  at the steady state between the wave modes can become significant under resonance conditions.

\subsection{Regular perturbation}\label{sec:dless}

We consider the case of bottom corrugations that are small compared to the mean waterdepth, i.e. we let $\epsilon=\f{d}{H}\ll 1$, such that the problem can be solved via asymptotic analysis. To ease the derivation, we use primed dimensionless variables given by $(u,v)=(Uu',Uv')$, $w=\mu U w'$, $b=d b'$, $\xh=\f{H}{\mu} \xh'$, $z=Hz'$, with $U$ the magnitude of the characteristic horizontal velocity of incident waves. The proposed scaling is suggested by the dimensional form of the wave modes over the flat bottom, i.e. equation \eqref{sc0}. With the proposed scaling, the dimensionless form of the flat-bottom modes reads
\ba{}\label{sc0bis}
w'_j=A'_j\sin \lb \kaj (z'+1)\rb e^{i\Kaj\cdot\xh'},
\ea
where $\Kaj=\f{\b{k}_jH}{\mu}$ is the dimensionless wavenumber ($ \kaj=|\Kaj|$), $w'_j \sim \sqrt{(u'_j)^2+(v'_j)^2} \sim \co(1)$, and the  bottom topography becomes
\ba{}\label{bi3bis}
b'=\f{1}{2} \lp e^{i\varphi}e^{i\Kab\cdot \xh'} + c.c. \rp,
\ea
with $\Kab=\f{\kb H}{\mu}$.% Note that unprimed variables will denote dimensionless variables from here onward, unless stated otherwise.

Using dimensionless variables the steady-state equations are written as (dropping the primes from here onward)
\bsa{sc10}\label{sc10a}
 \lp \p_z^2-\nabla_h^2 \rp  w &= 0,~ &-1+\epsilon b \leq z \leq 0, \\\label{sc10b}
 w &=0,~&z=0, \\ \label{s10}
 w &= \epsilon \uh\cdot\nabla_h b,~ &z=-1+\epsilon b, 
\esa 
and the polarization relations become
\ba\label{poless2}
\nabla_h^2 u = -\lp\p_x+\f{if}{\omega}\p_y\rp\p_z w, ~
\nabla_h^2 v = -\lp\p_y-\f{if}{\omega}\p_x\rp\p_z w.
\ea
Let us approximate the solution by a perturbation series of the form 
\ba{}\label{sc12}
\b{u} = \b{u}^{(0)}+\epsilon^1\b{u}^{(1)}+...
\ea
with $\b{u}^{(j)}\sim \co(1)$ for $j=1,2$. Substituting \eqref{sc12} into equation \eqref{s10} yields
\ba{}\notag
w^{(0)}&+\epsilon w^{(1)}+...= \epsilon\lb u^{(0)}+\epsilon u^{(1)}+... \rb\p_x b + \epsilon\lb v^{(0)}+\epsilon v^{(1)}+... \rb\p_y b, ~z=-1+\epsilon b,
\ea
which, using a first-order Taylor series approximation about $z=-1$, can be further expanded as
\ba{}\notag
w^{(0)}&+\epsilon b \p_z w^{(0)}+\epsilon w^{(1)}+\epsilon^2 b \p_z w^{(1)}+... = \epsilon\lb u^{(0)}+\epsilon b \p_z u^{(0)}+\epsilon u^{(1)}+\epsilon^2 b \p_z u^{(1)}+... \rb\p_x b \\ \label{sc3}
&+\epsilon\lb v^{(0)}+\epsilon b \p_z v^{(0)}+\epsilon v^{(1)}+\epsilon^2 b \p_z v^{(1)}+... \rb\p_y b,  ~z=-1.
\ea
The zero- and first-order equations obtained from \eqref{sc3} become, respectively,
\bsa{bi4}\label{bi4b}
&w^{(0)}=0,~z=-1, \\ \label{bi5}
&w^{(1)}=u^{(0)}\p_x b+v^{(0)}\p_y b-b\p_zw^{(0)}=\nabla_h \cdot \lp \uh^{(0)}b \rp ,~z=-1,
\esa
where in \eqref{bi5} we used the fact that $\b{u}^{(0)}$ must be divergence free and we note $\uh^{(0)}=(u^{(0)},v^{(0)})$. Note that in addition to the bottom boundary conditions \eqref{bi4b}-\eqref{bi5}, $\b{u}^{(0)}$ and $\b{u}^{(1)}$ must satisfy the same dimensionless momentum equation, top boundary condition, and polarization relations given by \eqref{sc10a}-\eqref{sc10b} and \eqref{poless2}. 

The asymptotic expansion resulting in equations \eqref{bi4} for the bottom boundary condition is valid provided that $\epsilon=d/H\ll 1$ and the right-hand side of \eqref{bi5} remains order one. For monochromatic bottom corrugations given by \eqref{bi3bis}, this means that we require (i) $ \kab \sim \co(1)$ (from the terms $u^{(0)}\p_x b$ and $v^{(0)}\p_y b$ on the right-hand side of \eqref{bi5}), and (ii) $\ka \sim \co(1)$ where $\ka$ is the dimensionless vertical wavenumber of internal waves (from the term $b\p_zw^{(0)}$ on the right-hand side of \eqref{bi5}). The first condition (i) implies in terms of dimensional variables $ k_b d/\mu \sim \co(\epsilon) \ll 1$, i.e. the bottom slope $k_b d$ must be much smaller than the internal-wave slope $\mu$, such that (i) is a stronger requirement than the typical subcritical-slope assumption $k_b d \leq \mu$ \cite[cf.][]{Buhler2011}. The second condition (ii) implies in terms of dimensional variables $ \gamma d \sim \co(\epsilon) \ll 1$, i.e. the corrugation height much be much smaller than the vertical wavelength ($2\pi/\gamma$) of internal waves. Note that (ii) is always verified  for the incident mode-one wave since the vertical wavelength is half the waterdepth.  

It can be noted that in considering the linear form  \eqref{bi2} of the wave equations, we have neglected the advection of bottom-generated waves by the incident flow. When the incident flow corresponds to semidiurnal internal tides, which is the case of our interest, this simplification is  acceptable because velocities of the internal tides are in general of order $U\sim\co(1cm/s)$ or smaller \cite[e.g.][]{Zhao2016} such that the advective term is much smaller than the acceleration term, i.e. in dimensional form $(U\p_x/\p_t)\sim (Uk/\omega) \ll 1$, for wavelengths down to a few kilometers or smaller.

%%%%%%%%%%%%%%%%%%%%%%
\subsection{Chain resonance conditions}\label{sec:bi21b}
%%%%%%%%%%%%%%%%%%%%%% 

As expected, the zero-order solution is not affected by bottom corrugations, such that it can be written as a superposition of wave modes given by \eqref{sc0bis}. Using the dimensionless polarization relations \eqref{poless2}, the forcing term on the right-hand-side of equation \eqref{bi5} due to the interaction of a zero-order incident wave mode \eqref{sc0bis} with bottom corrugations can thus be written as
\ba{}\label{forcing}
i \lb (\cos\theta+\f{if}{\omega}\sin\theta) \p_x\lp be^{i\Kaj\cdot\xh}\rp + (\sin\theta-\f{if}{\omega}\cos\theta) \p_y\lp be^{i\Kaj\cdot\xh}\rp \rb,
\ea
where $\theta$ denotes the direction of propagation of the internal waves, i.e. we note $\Kaj=(\kaj\cos\theta,\kaj\sin\theta)$.

For incident waves with wavenumber $\Kaj=\Kain$ and $b$ given by  \eqref{bi3bis}, it can be seen from \eqref{forcing} that the forcing of $w^{(1)}$ is by wave-like terms with wavenumber $\Kain \pm \Kab$. Because $w^{(0)}$ and $w^{(1)}$ satisfy the same governing equations (except for the bottom boundary condition), we thus know from the theory of wave-bottom resonance that $w^{(1)}$ can grow unbounded over the corrugated patch and potentially become much larger than $\co(1)$ when $\Kain\pm \Kab$ is a natural wavenumber of the zero-order problem, i.e. when $\exists j$ such that $|\Kain\pm \Kab| = j\pi$ \cite[see e.g.][for the case of surface waves]{Liu1998}. Regular perturbation fails to capture an accurate solution at resonance (the unbounded growth of $w^{(1)}$ is unphysical) because the wave-bottom interaction are so strong that the incident wave loses significant energy and new waves become strongly and coherently generated by the many small seabed undulations. This resonant scattering can be dealt with by the method of multiple scales, i.e. including all possibly blowing-up modes at leading-order and allowing them to slowly evolve over the corrugations, as done in section \S\ref{sec:bi3}. It is important to note here that the wave-bottom resonance is a resonance condition on the horizontal wavenumbers only because the interaction happens at a boundary between a freely-propagating wave and a frozen bottom wave (i.e. of zero frequency). This is a situation different from nonlinear wave-wave interactions for which the resonance condition requires a frequency match and a wavenumber match in all spatial dimensions.  %We The dynamics of energy transfers between all the resonated modes can then be solved using multiple-scale analysis, see section \S \ref{sec:bi3}. 

\begin{figure}
\centering
\includegraphics[width=0.9\textwidth]{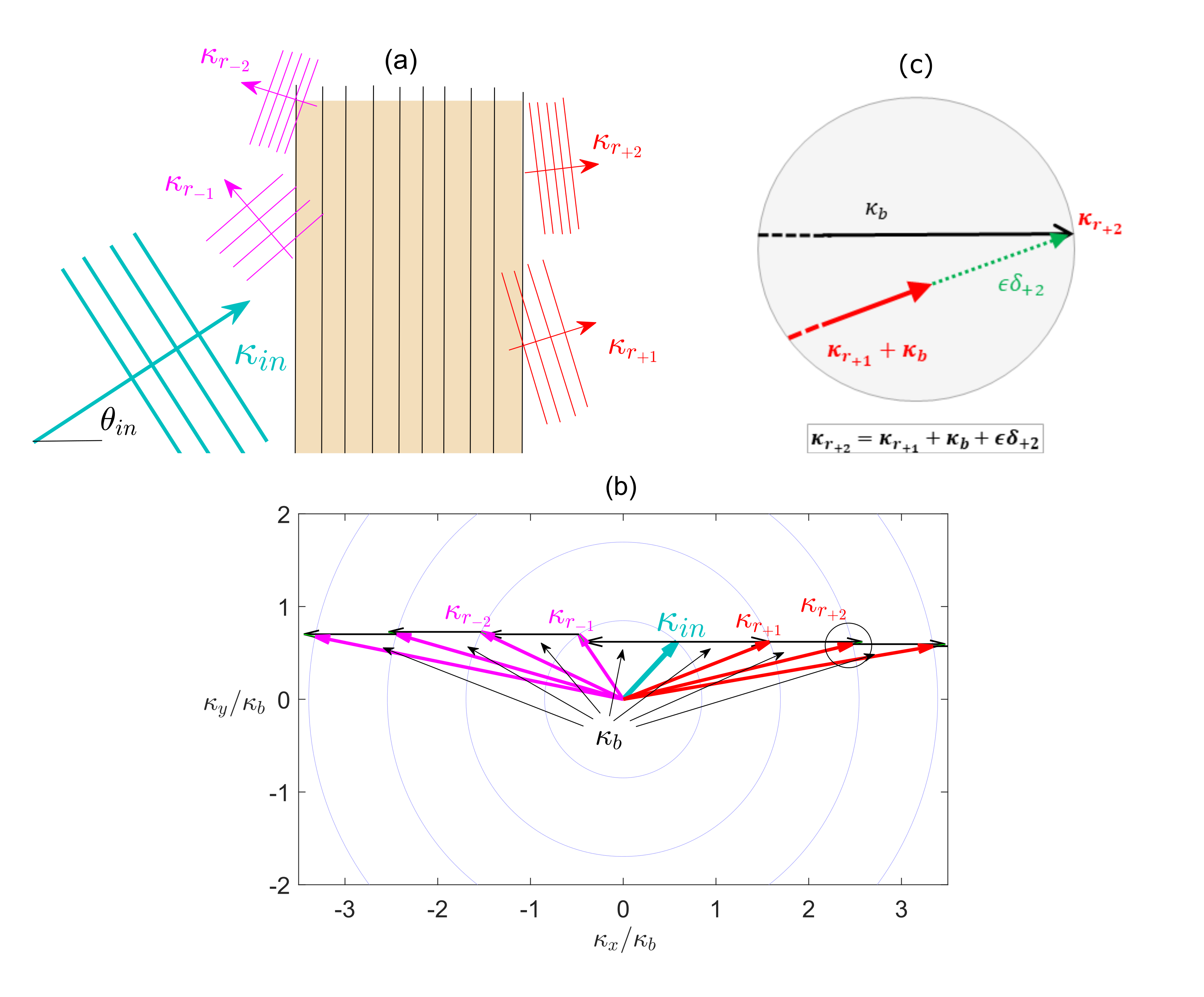}
\vspace{-.2in}
\caption{(a) Top view schematic of incident internal waves with dimensionless wavenumber $\kain$ and arbitrary incidence angle $\theta_{in}$, propagating toward a finite number of monochromatic seabed bars with which they can resonate new waves of varying wavelengths and directions of propagation. Figure (b) shows the chain resonance diagram in wavenumber space obtained when $\theta_{in}\sim\pi/4$. The light-colored concentric circles in the background represent the natural dimensionless wavenumbers of the system, i.e. solutions of the dispersion relation $\kappa_j=j\pi$ ($j$ an integer). Upon interacting resonantly with the seabed bars, the incident $\kain$ waves generate the transmitted $\kappa_{r_{+1}}$ and reflected $\kappa_{r_{-1}}$ waves, which themselves resonate the new waves $\kappa_{r_{\pm 2}}$. The process repeats itself with subsequent waves possibly infinitely. The successive resonances cannot be all perfect, however, which we clearly show in (c) by zooming in on the circle highlighting $\kappa_{r_{+2}}$ in (b): the resonant natural wavenumber $\bm{\kappa}_{r_{+2}}$ can be seen to be the result of the interaction wavenumber $\bm{\kappa}_{r_{+1}}+\kab$, plus a small detuning term $\epsilon\bm{\delta}_{+2}$.  }\label{resodiag}
\end{figure} 

When a resonance is obtained between the incident wave and the bottom corrugations, the newly (resonantly) generated wave is added to the set of zero-order wave modes and we note the corresponding wavenumber $\bm{\kappa}_{r_{\pm 1} }=\Kain\pm \Kab$ (cf. figure \ref{resodiag}a). In a homogeneous fluid, the interaction of an incident wave with monochromatic corrugations can resonate at most one wave (i.e. the reflected wave $\bm{\kappa}_{r_{-1}}$). In a continuously-stratified fluid, however, the initial interaction can resonate both $\bm{\kappa}_{r_+1}$ and $\bm{\kappa}_{r_{-1}}$, and each one of these two waves can then resonate other new waves. Such a chain resonance is shown in figure \ref{resodiag}b, which includes the incident mode-one internal wave ($\Kain$) interacting with monochromatic corrugations ($\Kab$), the resulting forcing (resonant) wavenumbers (noted $\bm{\kappa}_{r_{\pm 1}}$), as well as the forcing (again resonant) wavenumbers $\bm{\kappa}_{r_{+ 2}}=\bm{\kappa}_{r_{+1}}+\Kab$ due to the interaction of $\bm{\kappa}_{r_{+ 1}}$ with the same monochromatic corrugations. Other forcing resonant higher-mode wavenumbers ($\bm{\kappa}_{r_{-1}},\bm{\kappa}_{r_{-2}},...$, $\bm{\kappa}_{r_{+2}},\bm{\kappa}_{r_{+3}},..$) also participate in the chain resonance, with the sequence being potentially infinite.    

As can be expected from a purely geometric argument, the successive forcing wavenumbers cannot be all perfectly resonant unless the incident wave comes at normal incidence (i.e. $\theta_{in}=0$), since in this case the spacing between the successively resonated wavenumbers is always $\kappa_1=\pi$ in a linearly-stratified fluid. In figure \ref{resodiag}c, we illustrate the effect of detuning due to oblique incidence by showing that the $\bm{\kappa}_{r_{+1}}$ wave is in fact only near resonance with the corrugations $\Kab$, such that the forcing and resonant wavenumbers, $\bm{\kappa}_{r_{+1}}+\Kab$ and $\bm{\kappa}_{r_{+2}}$ respectively, are separated by a (small) detuning wavenumber, which we note $\epsilon\bm{\delta}_{+2}$. 

In the general case, we write the positive and negative branches of the successively-resonated wavenumbers as
\bsa{bi9}\label{bi91}
\bm{\kappa}_{r_{+m}}=\bm{\kappa}_{r_{+(m-1)}}+\Kab+\epsilon\bm{\delta}_{+m}, \\ \label{bi92}
\bm{\kappa}_{r_{-m}}=\bm{\kappa}_{r_{-(m-1)}}-\Kab+\epsilon\bm{\delta}_{-m},
\esa
for $m \geq 1$, with $\bm{\kappa}_{r_{\pm 0}}=\Kain$ the incident wavenumber. The resonated wavenumbers $\bm{\kappa}_{r_{\pm m}}$ are  obtained such that the detuning $\epsilon\|\bm{\delta}_{\pm m}\|_2=\epsilon\dm$ is minimum (i.e. using the $L^2$-norm). A chain resonance is expected provided that there is a number of successive detuning wavenumbers $\epsilon \dm$ that are small compared to $\kappa_1=\pi$, which is the wavenumber spacing between any two neighbouring natural wavenumbers. For the case of figure \ref{resodiag}, the detuning wavenumbers $\epsilon\{\delta_{-4},\delta_{-3},\delta_{-2},\delta_{-1},\delta_{+1},\delta_{+2},\delta_{+3}\}$ are all much smaller than $\co(1)$, such that the chain resonance includes at least 7 resonated waves. It should be noted that the wavenumber $\bm{\kappa}_{r_{+m}}$ is not \textit{a priori} an $m^{th}$ wavenumber mode, which is why we use the cumbersome subscripts $r_{\pm m}$ with $r$ standing for resonant, $\pm$ the positive or negative branch of the resonance, and $m$ the interaction number.

%%%%%%%%%%%%%%%%%%%%%%
\section{Derivation of the envelope equations}\label{sec:bi3}
%%%%%%%%%%%%%%%%%%%%%%

In order to obtain a uniformly-valid solution of energy exchanges between the incident and bottom-generated waves at resonance, we now introduce the slow variable $\sxh=\epsilon\xh$ and look for a solution $\b{u}(\xh,\sxh;\epsilon)$ that depends on both $\xh$ and $\sxh$. 

Inserting the multiple-scale ansatz $\b{u} = \b{u}^{(0)}(\xh,\sxh;\epsilon)+\epsilon^1\b{u}^{(1)}(\xh,\sxh;\epsilon)+...$ into equations \eqref{sc10} yields at zero order
\bsa{new1}\label{new1a}
\lp \p_z^2-\nabla_h^2 \rp w^{(0)} & = 0,~&-1\leq z \leq 0, \\ \label{new1b}
w^{(0)} & = 0,~&z=0, \\ \label{new1c}
w^{(0)} & =0,~ &z=-1.
\esa
Equation \eqref{new1} is the standard equation for internal waves over a flat bottom, which means that the fast variations of $w^{(0)}$ can be simply written as a sum of the same internal wave modes as in \eqref{sc0bis}. The fast variations of $w^{(0)}$ are decoupled from the slow variations, which means that we can use a separation of fast/slow variables. This leads us to consider the general expression for $w^{(0)}$:  
\ba{}\label{new2}
w^{(0)}=  A_{in}(\sxh)S_{in}(z)E_{in}(\xh) + \sum_{m=1}^{\infty} A_+(\sxh)S_{+ m}(z)E_{+ m}(\xh) +  \sum_{m=1}^{\infty} A_-(\sxh)S_{- m}(z)E_{- m}(\xh),
\ea
where $ A_{in}(\sxh)$ and $\Am(\sxh)$ are the slowly-varying amplitudes of the incident and resonated waves to be found, $\Kam$ are the resonated wavenumbers given by the chain resonance equations \eqref{bi9}, $S_{\pm m} = \sin [\kam(z+1)]$ is the zero-order sine vertical structure function (recall $\kam=\pi\rm$ with $\rm$ the wave-mode number), and $E_{\pm m}=\exp(i\Kam \cdot\xh)$. From the dimensionless steady-state polarization relations \eqref{poless2} we find at zero order
\bsa{uv}
u^{(0)}=\sum_{m=0}^{\infty} i\Am(\sxh) \lb \cos(\tm)+\f{if}{\omega}\sin(\tm)\rb C_{\pm m}(z)E_{\pm m}(\xh), \\
v^{(0)}=\sum_{m=0}^{\infty} i\Am(\sxh) \lb \sin(\tm)-\f{if}{\omega}\cos(\tm)\rb C_{\pm m}(z)E_{\pm m}(\xh),
\esa
where $m=0$ denotes the incident wave mode, $C_{\pm m} = \cos [\kam(z+1)]$ is the zero-order vertical cosine structure functions, and we note $\theta_{in}$ and $\tm$ the angles of the incident and resonated wavenumbers $\Kain$ and $\Kam$ with the $x$ axis.

Inserting the multiple-scale ansatz $\b{u} = \b{u}^{(0)}(\xh,\sxh;\epsilon)+\epsilon^1\b{u}^{(1)}(\xh,\sxh;\epsilon)$ into equations \eqref{sc10} yields at first order
\bsa{bi13}\label{bi131}
\lp \p_z^2-\nabla_h^2 \rp w^{(1)} & = 2 \snab\cdot\nabla_h w^{(0)},~&-1\leq z \leq 0, \\ \label{bi132}
w^{(1)} & = 0,~&z=0, \\ \label{bi133}
w^{(1)} & = \nabla_h \cdot\lp \uh^{(0)}b \rp,~ &z=-1,
\esa
with $\snab=(\p_{\x},\p_{\y})$. The right-hand sides of equations \eqref{bi131} and \eqref{bi133} are the result of the slow variations of the wave amplitudes $\Am$ of the zero-order solution $w^{(0)}$, and of the zero-order-wave--bottom interactions (as in regular perturbation). Thus the multiple-scale solution is valid provided that variations of $\Am$ are slow and the assumptions of regular perturbation are verified (cf. \S\ref{sec:dless}). Because the homogeneous first-order problem is the same as the zero-order problem, the forcing terms on the right-hand sides of equations \eqref{bi13} can be solutions (eigenmodes) of the left-hand-side (homogeneous) operator. When the forcing is an eigenmode, the forcing is said to be resonant. Unlike regular perturbation, however, a bounded solution $w^{(1)}$ can still be obtained at resonance provided that the forcing terms satisfy a solvability condition \cite[][]{Fredholm1903}, which corresponds to an equation for the zero-order waves on the slow scale. 

The solvability condition ensures that $w^{(1)}$ remains bounded by requiring the resonant forcing due to the slow variations of the wave amplitudes (represented by the right-hand-side of equation \eqref{bi131}) to counterbalance the resonant forcing due to the wave-bottom interactions (equation \eqref{bi133}). Mathematically the solvability condition is obtained as the condition that the direct integration of the forcing \eqref{bi131} projected onto the eigenmodes $S_{\pm m}E^*_{\pm m}$ of the linear operator \cite[e.g.][]{Mei1985}
\ba{}\notag 
&\int\text{d}x\text{d}y\int_{z=-1}^{0}\text{d}z \lb S_{\pm m}E^*_{\pm m} \lp \p_z^2-\nabla_h^2 \rp  w^{(1)} \rb  \\  \notag
= &\int\text{d}x\text{d}y   E^*_{\pm m} 2 \snab\cdot\nabla_h\int_{z=-1}^{0}\text{d}z S_{\pm m} w^{(0)} , \\   \label{cond1}
= &\int\text{d}x\text{d}y E^*_{\pm m} \snab\cdot\nabla_h \lp \sum_{\ell=0}^{\infty} A_{\pm \ell} E_{\pm \ell} \rp , 
%= &ih(\Km\cdot\bar{\nabla}_H) \Am,
\ea
derived with the help of \eqref{bi131} ($^{*}$ denotes the complex conjugate), is equal to the integral value projected on the boundaries (obtained following Green's method), i.e. 
\ba{}\notag
&\int\text{d}x\text{d}y\int_{z=-1}^{0} \text{d}z \lb S_{\pm m}E^*_{\pm m} \lp \p_z^2-\nabla_h^2 \rp  w^{(1)} \rb \\  \notag
= &\int\text{d}x\text{d}y\int_{z=-1}^{0} \text{d}z \lb S_{\pm m}E^*_{\pm m} \lp \p_z^2-\nabla_h^2 \rp  w^{(1)} - w^{(1)} \lp \p_z^2-\nabla_h^2 \rp S_{\pm m}E^*_{\pm m} \rb, \\\notag
= &\int\text{d}x\text{d}y \lcb \lb S_{\pm m}\p_z w^{(1)}\rb_{z=-1}^{0}E^*_{\pm m}-\lb w^{(1)} \p_z S_{\pm m}\rb_{z=-1}^{0}E^*_{\pm m}\rcb  , \\ \label{cond2}
= &\int\text{d}x\text{d}y\kam \lcb  \underbrace{\p_x \lb u^{(0)}(z=-1)b \rb}_{\mathcal{X}} + \underbrace{\p_y \lb v^{(0)}(z=-1)b \rb}_{\mathcal{Y}} \rcb E^*_{\pm m},
\ea
where, using the chain resonance equations for the resonated wavenumbers \eqref{bi9}, 
\ba{}\notag
\mathcal{X} & = \p_x \lcb \f{1}{2} \lp e^{i\varphi} e^{i\Kab\cdot \xh} + e^{-i\varphi} e^{-i\Kab\cdot \xh} \rp  \sum_{n=0}^{\infty} i\An \lb \cos(\tn)+\f{if}{\omega}\sin(\tn)\rb E_{\pm n} \rcb \\ \notag
& = \p_x \lcb \f{1}{2}  \sum_{n=0}^{\infty} i\An \lb \cos(\tn)+\f{if}{\omega}\sin(\tn)\rb \lp e^{ i\varphi} e^{i(\Kan + \Kab )\cdot \xh} + e^{ -i\varphi} e^{i(\Kan - \Kab )\cdot \xh}\rp  \rcb \\ \notag
& = \p_x \lcb \f{1}{2}  \sum_{n=0}^{\infty} iA_{+n} \lb \cos(\theta_{+n})+\f{if}{\omega}\sin(\theta_{+n})\rb \lp e^{ i\varphi} e^{i\bm{\kappa}_{r_{+ (n+1)}}\cdot \xh}e^{-i\bm{\delta}_{+(n+1)}\cdot \sxh} \right. \right. \\ \notag
& \left. \left. + e^{ -i\varphi} e^{i\bm{\kappa}_{r_{+ (n-1)}}\cdot \xh}e^{i\bm{\delta}_{+n}\cdot \sxh} \rp + \f{1}{2}\sum_{n=0}^{\infty} iA_{-n} \lb \cos(\theta_{-n})+\f{if}{\omega}\sin(\theta_{-n})\rb \right. \\ \notag
& \left. \lp e^{ i\varphi} e^{i\bm{\kappa}_{r_{-(n-1)}}\cdot \xh}e^{i\bm{\delta}_{-n}\cdot \sxh} + e^{ -i\varphi} e^{i\bm{\kappa}_{r_{- (n+1)}}\cdot \xh}e^{-i\bm{\delta}_{-(n+1)}\cdot \sxh} \rp \rcb,
\ea
and where the expression for $\mathcal{Y}$ is the same as $\mathcal{X}$ provided that we make the changes $\p_x\rightarrow\p_y$ and $f\rightarrow -f$.

%
%yielding
%
%\ba{}\label{solva}
%ih(\Km\cdot\bar{\nabla}_H) \Am  = \f{\int_{-\infty}^{\infty} \text{d}x \lcb E^*_{\pm m}\f{\gm\mu^2}{\epsilon} \lb \p_x \lp u^{(0)}b \rp + \p_y \lp v^{(0)}b \rp \rb{\biggr\rvert_{z=-h}}\rcb}{ \int_{-\infty}^{\infty} \text{d}x }.
%\ea
%

The equation for each wave amplitude $A_{\pm m}(\sxh)$ is  obtained by equating nonzero terms in equations \eqref{cond1} with those in \eqref{cond2}  for all $m$ ($\int\text{d}x\text{d}y \equiv 0$ unless the integrand is nonoscillatory). In equation \eqref{cond1}, nonzero terms are trivially given by $i\Kam\cdot\snab\Am$. In equation \eqref{cond2}, nonzero terms are obtained by selecting the wave modes $A_{\pm(m-1)}$ and $A_{\pm(m+1)}$ within the sums in $\mathcal{X}$ and $\mathcal{Y}$. Equating the nonzero terms, the equation for $\Am$ is finally obtained as
\ba{}\notag
&\lb \lp \f{\Kam}{\kam} \rp \cdot \snab \rb \Am = \f{i\kam}{2} \lb \lp \f{\Kam}{\kam} \rp \cdot + \f{if}{\omega} \lp \f{\Kam}{\kam} \rp \times \rb   \\ \label{bi18} 
&\lcb \lp \f{\bm{\kappa}_{r_{\pm(m-1)}}}{\kappa_{r_{\pm(m-1)}}} \rp A_{\pm(m-1)}  e^{-i\Dm\cdot\sxh}e^{\pm i\varphi} + \lp \f{\bm{\kappa}_{r_{\pm(m+1)}}}{\kappa_{r_{\pm(m+1)}}} \rp A_{\pm(m+1)} e^{i\bm{\delta}_{\pm(m+1)}\cdot\sxh} e^{\mp i\varphi} \rcb,
\ea
where we recall that $\dm$ is the detuning wavenumber (cf. equation \eqref{bi9}) which must be order $\co(1)$ for the multiple-scale derivation to be valid, and ``$\cdot$'' and ``$\times$''  (inside the brackets on the right-hand side of \eqref{bi18}) are the dot and cross product operators acting on the (vector) terms inside the curly brackets. The left-hand-side of equation \eqref{bi18} represents the slow variations of $\Am$, while the right-hand-side represents the effect of the interactions between the corrugations and its resonant neighbors, i.e. $A_{\pm(m-1)}$ and $A_{\pm(m+1)}$. The strength of energy exchanges between the two coupled pairs of waves is, for the case of colinear waves, given by $\kam$. Thus, energy exchanges occur over a shorter length scale if the waves' wavenumber $\kam$ is larger. Provided that there is no detuning (i.e. $\delta_{\pm m}=0$), energy exchanges also occur over a shorter length scale if the waves propagate in the same direction (from the $\Kam\cdot \bm{\kappa}_{r_{\pm(m\pm 1)}}$ terms), since in this case the imaginary (weakening) Coriolis term is 0. With nonzero detuning, however, the effect of Coriolis becomes more complex and no general conclusion can be drawn. It is interesting to note that the exchange of energy  through wave-bottom interactions between waves that are perpendicular ($\Kam\cdot\bm{\kappa}_{r_{\pm (m\pm 1)}}=0$), which is impossible in a nonrotating system (a phenomenon known as Bragg degenerate resonance in the case of surface waves, see e.g. \cite{Couston2017}), becomes possible with the addition of the Coriolis effect. A straightforward manipulation and combination of equations \eqref{bi18} (i.e. multiplying by $A_{\pm m}^*/\kam$ with $^*$ denoting the complex conjugate, adding the complex conjugate counterpart, and summing all equations over $\pm m$) can be shown to result in an equation for the conservation of total energy flux, i.e.
\ba{}\label{flux}
\snab\cdot \lp \f{\Kain}{\kain^2}|A_{in}|^2+\sum_{m=1}^{\infty} 
\f{\bm{\kappa}_{r_{+m}}}{\kappa^2_{r_{+m}}}|A_{+m}|^2 +\sum_{m=1}^{\infty} 
\f{\bm{\kappa}_{r_{-m}}}{\kappa^2_{r_{-m}}}|A_{-m}|^2 \rp =0,
\ea
which is expected since the  system is conservative and at steady state (recall $A_0=A_{in}$). Note that since the dimensional horizontal group velocity is $\b{C}_g=(\omega^2-f^2)\b{k}/[\omega(1+\mu^2)k^2]$ and $\omega$, $\mu$ are the same for all waves, the expression to the right of the gradient in equation \eqref{flux} is indeed the total horizontal energy flux. 

The envelope equation can be rewritten in a more convenient form, i.e. with constant coefficients,  provided that we consider the modulated version of the envelope amplitude
\ba{}\label{modulated}
\Am\rightarrow\tilde{A}_{\pm m}e^{-i\D2m\cdot\sxh},
\ea
where
\ba{}\label{bi25}
\D2m=\bm{\delta}_{\pm 1}+...+\Dm.
\ea
Substituting \eqref{modulated} in equation \eqref{bi18} yields, dropping the tilde symbol,
\ba{}\notag
&\lb \lp \f{\Kam}{\kam} \rp \cdot \snab \rb \Am = \f{i\kam}{2} \lb \lp \f{\Kam}{\kam} \rp \cdot + \f{if}{\omega} \lp \f{\Kam}{\kam} \rp \times \rb   \\ \label{bi40} 
&\lcb \lp \f{\bm{\kappa}_{r_{\pm(m-1)}}}{\kappa_{r_{\pm(m-1)}}} \rp A_{\pm(m-1)} e^{\pm i\varphi} + \lp \f{\bm{\kappa}_{r_{\pm(m+1)}}}{\kappa_{r_{\pm(m+1)}}} \rp A_{\pm(m+1)}e^{\mp i\varphi}  \rcb +i \f{(\Kam\cdot\D2m)}{\kam\cos\tm}A_{\pm m},
\ea
which is a system of ordinary-differential equations with constant coefficients  easily amenable to numerical treatment. Physically, the energy exchanges between $A_{\pm m}$ and its resonant neighbors occur after the successive (detuned) transfers of energy from the incident to the $A_{\pm m}$ wave, such that energy exchanges are subject to the cumulative detuning of the chain resonance rather than the individual pair detunings. This explains the presence of the cumulative detuning term $\D2m$ on the righ-hand-side of \eqref{bi40}. Note that for the positive (resp. negative) branch of the chain resonance, we take in equation \eqref{bi40}  the upper $+$ (resp. lower  $-$) sign of the subscript indices.

In the results section (\S\ref{sec:bi4}) we restrict our attention to sinusoidal corrugations that are long-crested and parallel to the $y$ axis, i.e. we set $\varphi=-\pi/2$ and assume $\Kab=\kab\hat{x}$ ($\hat{x}$ denoting the unit vector in $x$ direction), such that $\p_y A_{\pm m}\equiv 0$. The envelope equation \eqref{bi40} thus becomes one dimensional and is simplified as (substituting the slow variable $\bar{x}$ with $\epsilon x=\f{d}{H}x$)
\ba{}\notag
&\f{\text{d} A_{\pm m}}{\text{d}x}  = \pm \f{d\kam}{2H} \lb A_{\pm(m-1)} \f{\cos(\tm-\theta_{\pm(m-1)})+\f{if}{\omega}\sin(\tm-\theta_{\pm(m-1)})}{\cos\tm} \right. \\ \label{bi19}
&\left. - A_{\pm(m+1)} \f{\cos(\tm-\theta_{\pm(m+1)})+\f{if}{\omega}\sin(\tm-\theta_{\pm(m+1)})}{\cos\tm} \rb 
+ i \f{(\Kam\cdot\f{d}{H}\D2m)}{\kam\cos\tm}A_{\pm m},
\ea
for the resonated waves $m\geq 1$ (note that the waves still propagate in both $x$ and $y$ even though slow variations only occur in $x$, as shown in figure \ref{resodiag}), and 
\ba{}\notag
 \f{\text{d} A_{in} }{\text{d}x} = \f{d\kain}{2H}  \lb A_{-1} \f{\cos(\theta_{in}-\theta_{-1})+\f{if}{\omega}\sin(\theta_{in}-\theta_{-1})}{\cos\theta_{in}} \right. \\ \label{bi19b}
\left. - A_{+1} \f{\cos(\theta_{in}-\theta_{+1})+\f{if}{\omega}\sin(\theta_{in}-\theta_{+1})}{\cos\theta_{in}} \rb,
\ea
for the incident wave ($m=0$). We solve equations \eqref{bi19}-\eqref{bi19b} as an eigenvalue problem over the domain $x\in[0,~\mathcal{N}\lambda_b]$ ($\mathcal{N}$ the number of corrugations and $\lambda_b=2\pi/\kab$ the dimensionless corrugation wavelength) with $x=0$ (resp. $x=\mathcal{N}\lambda_b$) the beginning (end) of the corrugated patch. The boundary conditions are $A_{in}=1$ at $x=0$, i.e. the incident wave has a constant unitary amplitude upstream of the bars, $A_{+m}=0$ at $x=0$, i.e. waves propagating in $+x$ direction grow in amplitude from $x=0$ onward, and $A_{-m}=0$ at $x=\mathcal{N}\lambda_b$, i.e. waves propagating in $-x$ direction can only exist upstream of the last seabed bar. As we solve equations \eqref{bi19}-\eqref{bi19b}, we consider a solution of the form
\ba{}\label{trunc}
w^{(0)}_{M^{\pm}}=  A_{in}S_{in}E_{in} + \sum_{m=1}^{M^+} A_{+m} S_{+ m}E_{+ m} + \sum_{m=1}^{M^-} A_{-m} S_{- m}E_{- m},
\ea
i.e. we consider a finite number $M^+$ and $M^-$ of resonated waves on both the positive and negative resonance branch (cf. equation \eqref{bi9}). We choose $M^+$ and $M^-$ such that increasing them (i.e. adding more modes to the solution) does not change the solution amplitudes of the lower wavenumber modes significantly, i.e. the amplitudes do not change more than $\sim 10\%$. The convergence of the truncated series \eqref{trunc} with $M^{\pm}$ toward the infinite-series solution \eqref{new2} is expected because high mode-number waves grow more slowly, i.e. they require a longer corrugated patch, than low mode-number waves (which besides carry most of the energy). It should be remarked that waves with relatively large detuning are not accurately predicted by multiple-scale analysis (since the derivation requires $\epsilon\dm\ll \kappa_1 = \pi$), but at the same time are not expected to carry a significant amount of energy as they result from wave-bottom interactions far from resonance. Therefore, here we check that waves with large detuning (i.e., specifically, $\epsilon\dm/\kappa_1 \geq 0.1$) included in \eqref{trunc} have small amplitudes (i.e. we check that the combined amplitude of detuned waves is always less than $<0.1$), such that their effect on the multiple-scale solution can be neglected.

%%%%%%%%%%%%%%%%%%%%%%
\section{Results for idealized oceanic conditions}\label{sec:bi4}
%%%%%%%%%%%%%%%%%%%%%%

Here we demonstrate through different case studies that the interaction of an incident internal wave with long-crested bottom corrugations can lead to the resonant generation of many new high-wavenumber internal waves, thus potentially contributing to enhanced mixing events in the oceans. The variations of the incident and resonated wave amplitudes are obtained from equations \eqref{bi19}-\eqref{bi19b} and the sequence of resonated wavenumber vectors is given by \eqref{bi9}. The solution is fully determined by specifying the incident wave parameters $\{\kain,\theta_{in},f/\omega\}$, the number of corrugations $\mathcal{N}$ and their amplitude $d/H$, and the corrugation wavenumber $\kab$ or alternatively the initial detuning $\delta_{\pm 1}$. 

We consider incident waves that are mode-one internal waves (i.e. $\kain=\kappa_1=\pi$) because such waves are known to carry significant amount of energy, to propagate long distances, and to interact with seabed topographies in the oceans \cite[][]{Zhao2010}. Mode-one internal waves are most often generated by surface tides or geostrophic flows over seabed topographies, and thus typically have low frequency, i.e. close to the tidal or inertial frequency. This implies that the internal-wave slope considered are relatively small, i.e. e.g. $\mu\sim 0.07-0.18$ for $N/f\sim 10-15$ and $\omega/f\sim 1.5-2$, and that the mode-one internal wavelength is in the $33-114$ km range for a mean water-depth $H\sim 3-4$ km. More specifically, here we consider incident waves at semi-diurnal tidal frequency $\omega=1.4\times 10^{-4} rad/s$ and Coriolis frequency $f=8.3\times 10^{-5} rad/s$ (obtained at $35\degree$ latitude), such that $f/\omega = 0.6$. It can be noted that for the case of long-crested corrugations parallel to the $y$ axis (as in figure \ref{resodiag}), the chain resonance tends to resonate waves going primarily in the $x$ direction, such that the Coriolis effect is in general weak. The efficiency of energy transfers becomes independent of the incident-wave frequency or slope, and hence of all wave parameters, when the incident wavenumber and corrugation wavenumber are colinear and/or $f=0$ \cite[][]{Buhler2011}.

Topographical features in the real ocean that are prominent and appear almost monochromatic are canyons and crests of fracture zones in the deep ocean. Such submarine features can be observed worldwide and in particular along the mid oceanic ridges in regions of the East Pacific Rise and Mid-Atlantic Ridge \cite[][]{StLaurent2002}, the Drake passage region in
the Southern Ocean \cite[][]{Nikurashin2010a}, and in the northeast Pacific \cite[][]{Bell1975}. The wavelength of submarine canyons and crests is generally in the $50-120$ km range, hence is similar to the wavelength of mode-one internal waves. As a result, here we consider small corrugation wavenumbers, i.e. $\kab\sim \kain=\kappa_1=\pi$, such that the first resonated waves are low-mode internal waves (i.e. 1 or 2). 

Because the multiple-scale theory requires $\ka \f{d}{H} \ll 1$ (cf. equation \eqref{bi4}), the normalized corrugation amplitude $d/H$ must be small compared to the inverse of the largest mode number obtained in the chain resonance. For chain resonance including up to 10 modes, for instance, the theory thus requires $d/H\sim \co(0.01)$. With such small corrugation amplitudes, a relatively large number of corrugations is required to obtain substantial energy transfers from the incident to the resonated waves. Therefore, here we will fix the number of corrugations to $\mathcal{N}=25$ for all our results. Much fewer seabed bars would be necessary if they were of a larger amplitude. However, satisfying the assumption $\ka \f{d}{H} \ll 1$ shall prevail, so we restrict our analysis to small seabed bars. Note that we made sure that the right-hand-side of equation \eqref{s10} is $\co(0.1)$ and always smaller than $1/3$ for all our results.

\subsection{Chain resonance at normal incidence}\label{ss1}

In this section we describe the chain resonance obtained at normal incidence and for long corrugations (i.e. $\kab\sim \kain=\pi$ and $\theta_{in}=0$). The incident wave can be either perfectly-tuned or detuned, i.e. such that $\epsilon\delta_{+1}=0$ or $\epsilon\delta_{+1}\neq 0$ (cf. \eqref{bi9}). We consider $\mathcal{N}=25$ seabed bars and the corrugation amplitude is $\epsilon=d/H=0.008$.

\begin{figure}
%\centering
\hspace{-0.05in}
\includegraphics[width=0.33\textwidth]{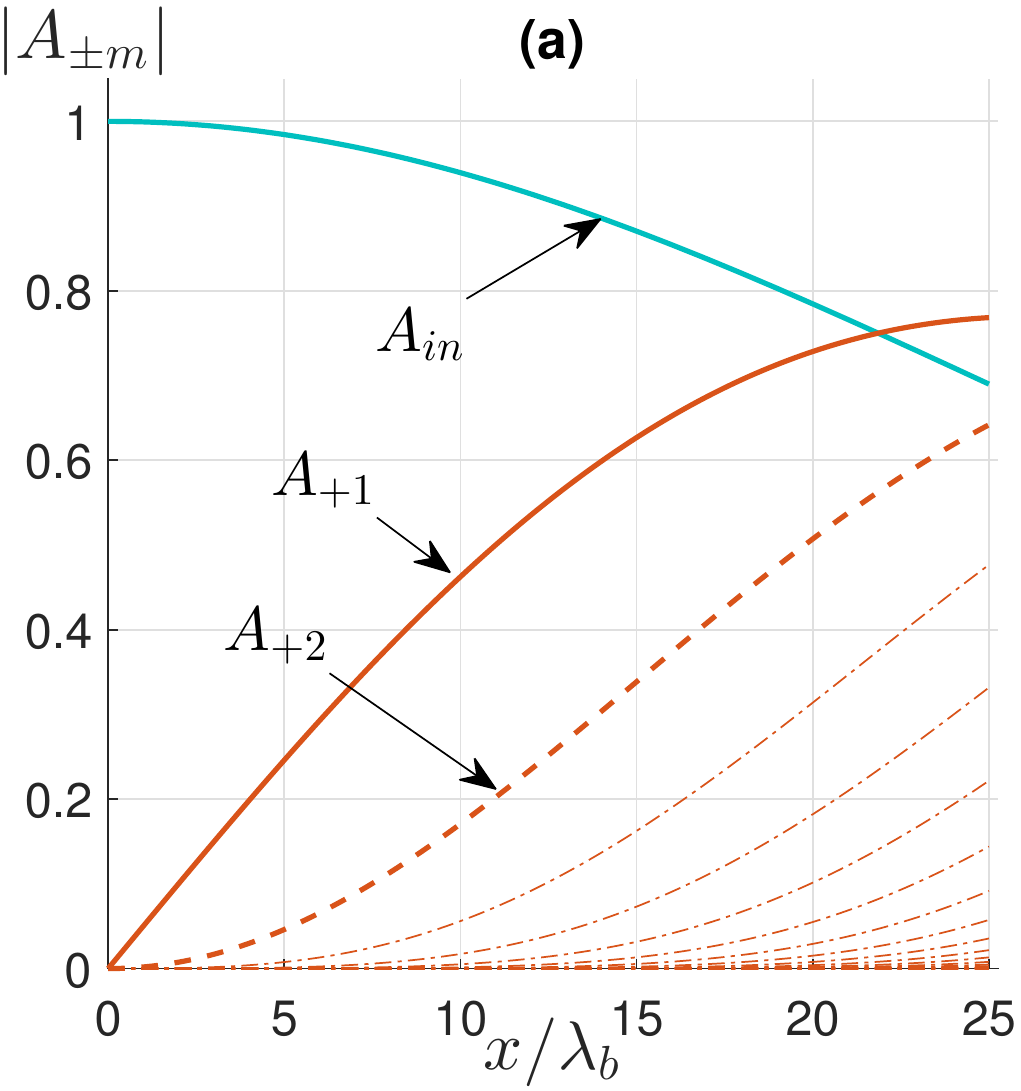}
\hspace{-0.in}\includegraphics[width=0.33\textwidth]{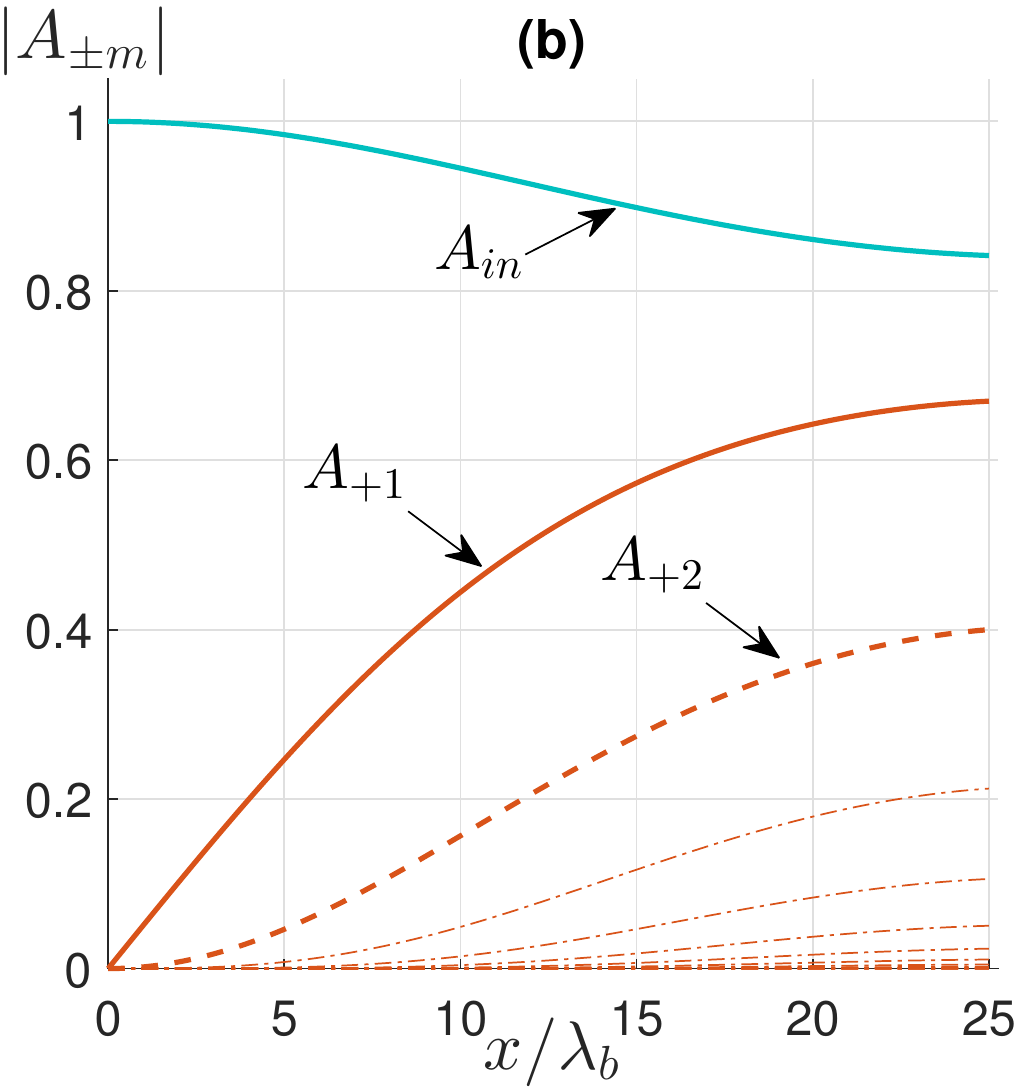}
\hspace{-0.0in}\includegraphics[width=0.33\textwidth]{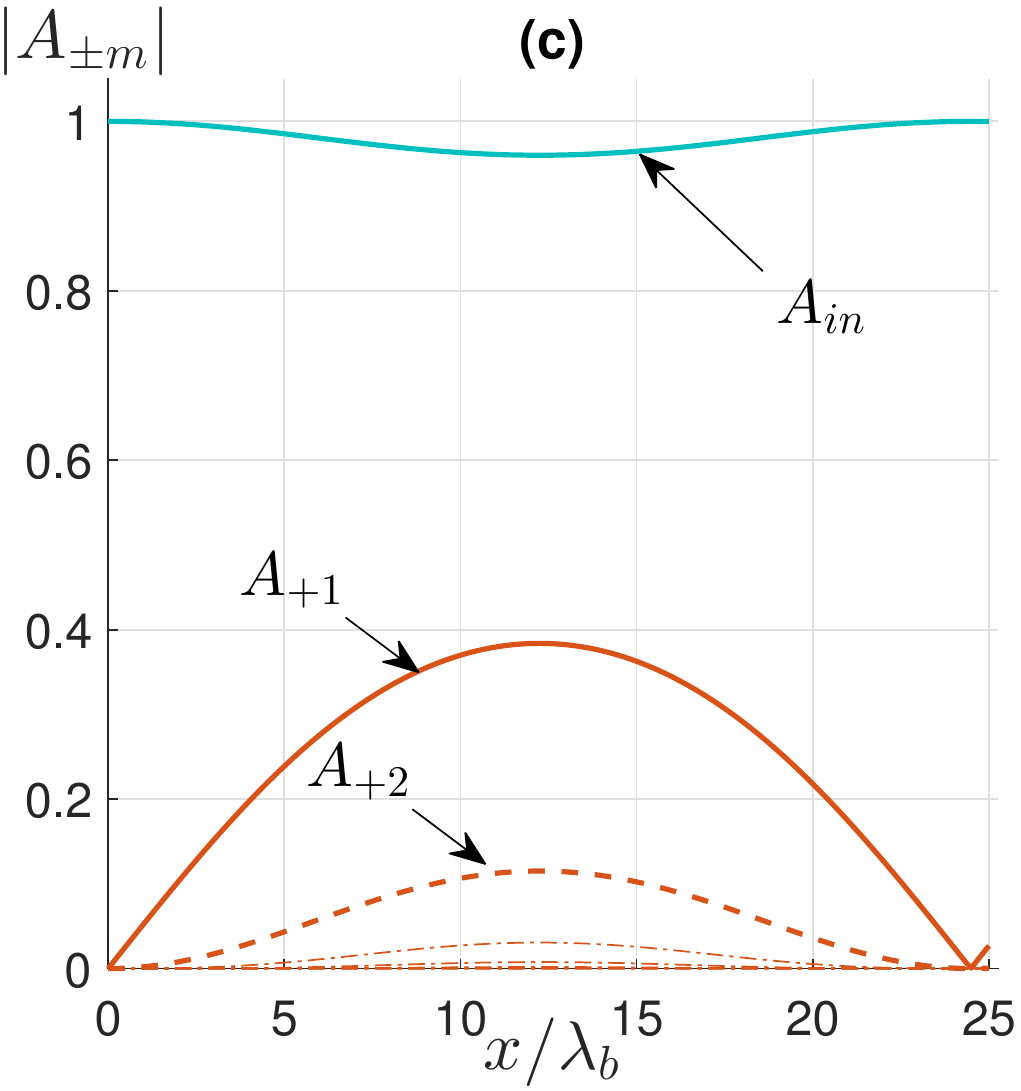}
\vspace{-0.15in}
\caption{Variations of the incident and resonated envelope amplitudes ($A_{in}$ and $A_{+1},A_{+2},...$) over 25 seabed bars for normal incidence ($\theta_{in}=0$, cf. figure \ref{resodiag}), as obtained from equations \eqref{bi19}-\eqref{bi19b} ($\lambda_b=2\pi/\kab$). With $\kab\sim \kain=\pi$, the first resonated wave $A_{+1}$ is a mode-two wave in all cases and there can be no back-scattered ($A_{-m}$) wave. The detuning of the incident wave $\epsilon\delta_{+1}/\kain$ is (a) 0, (b) 2\%, (c) 4\%. Increasing detuning results in fewer resonated waves and a re-capture of the available energy by the incident wave before the end of the patch [(c)]. The corrugation amplitude is $d/H=0.008$. Note that Coriolis has no effect on the variations of envelope amplitudes in this case of colinear waves.}\label{detfor}
\end{figure}

The perfectly-tuned one-dimensional case was briefly discussed in \cite{Buhler2011}, following an early derivation by \cite{Chen2009} who obtained a closed-form solution for the amplitude variations. Solving equations \eqref{bi19}-\eqref{bi19b} with $\kab=\kappa_1$, we find the variations of the incident ($A_{in}$) and resonated ($A_{+m}$) wave amplitudes as shown in figure \ref{detfor}a; there are no back-scattered $A_{-m}$ waves because the spacing between the corrugations is large and the detuning of the $\kappa_{r_{-1}}$ wave is also large. Note that our results agree perfectly with the solution of \cite{Chen2009}, which is expected since we solve \eqref{bi19}-\eqref{bi19b} with a sufficiently large number of waves (i.e. $M^{\pm}$ in \eqref{trunc} is sufficiently large that the truncated solution is close to the infinite-series solution). The chain resonance for perfectly-tuned normally-incident waves only includes perfect resonances, such that we observe a cascade of the incident wave energy from the longest to the smallest scales. The initial amplitude decays monotonically over the bars while newly resonated waves successively capture the incident wave energy.

When the first wave-bottom interaction is not perfectly resonant, we obtain the amplitude variations shown in figure \ref{detfor}b-c. For $\epsilon\delta_{+1}/\kain=0.02$, figure \ref{detfor}b shows that the energy cascade still takes place, albeit at a slower rate over the seabed bars compared to the perfectly-tuned case. With increased detuning $\epsilon\delta_{+1}/\kain=0.04$, however, we can see in figure \ref{detfor}c that the chain resonance is stopped at roughly the second or third resonated wave. We also observe that the incident wave even gains energy back from the resonated waves, such that downstream of the patch the incident mode-one internal wave is again close to 1. With half the number of seabed bars, the incident wave amplitude would on the contrary be at its minimum, as can be seen at $x/\lambda_b\sim 12.5$ in figure \ref{detfor}c. Indeed, the solution obtained for forward-scattered waves only is independent of the number of seabed bars, such that the results shown in figure \ref{detfor}c also apply for smaller patches. As the detuning is further increased, energy exchanges become even less strong and oscillate more rapidly. With $\epsilon\delta_{+1}/\kain= 0.1$, the decrease of the incident wave is negligible. It should be remarked, however, that detuning effects depend upon the interaction strength, measured as $d/H$. From equation \eqref{bi19}, it can be seen that the detuning term to the interaction term ratio is proportional to $(\epsilon\Delta_{\pm m}/\kam)/(d/H)$, such that detuning effects become less important as $d/H$ is increased.

\begin{figure}
%\centering
\hspace{-0.05in}
\includegraphics[width=0.33\textwidth]{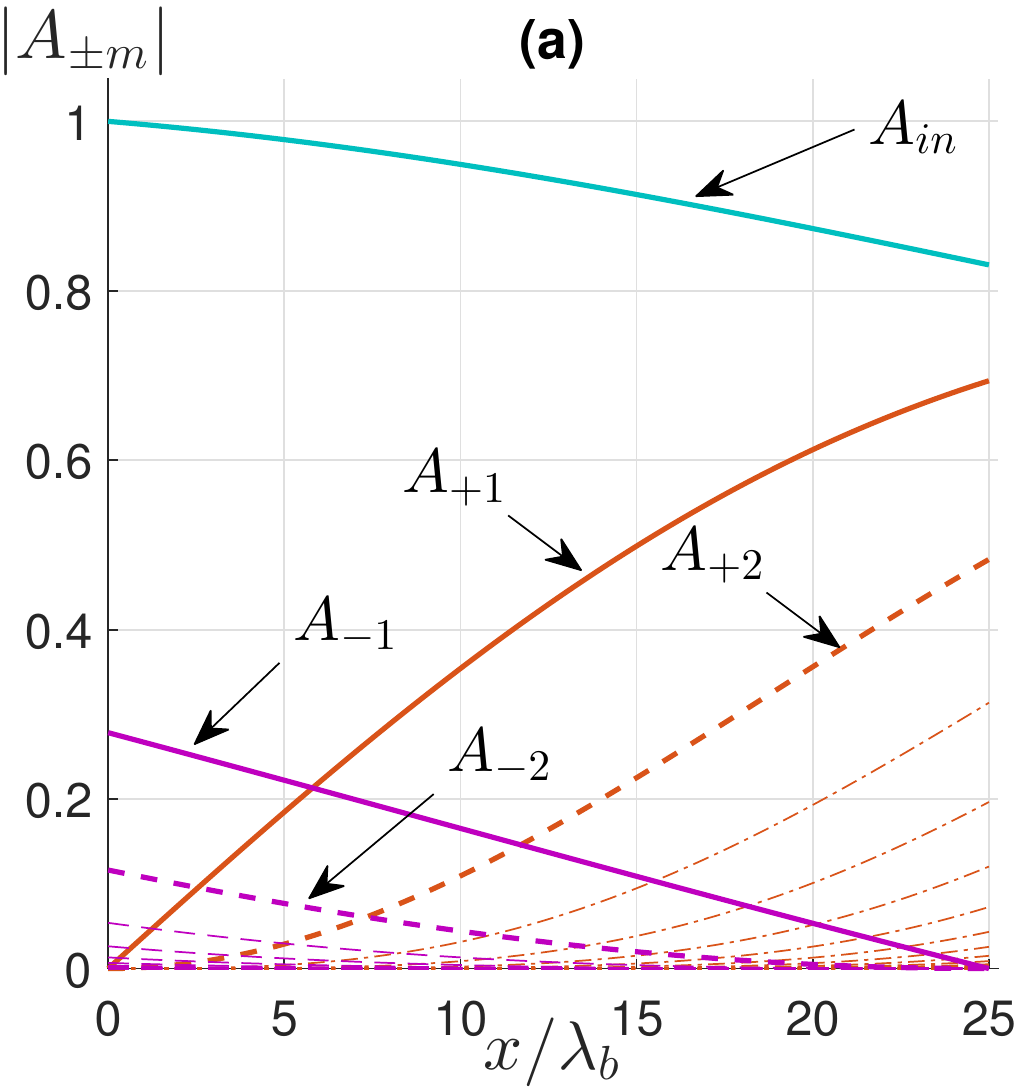}
\hspace{-0.in}\includegraphics[width=0.33\textwidth]{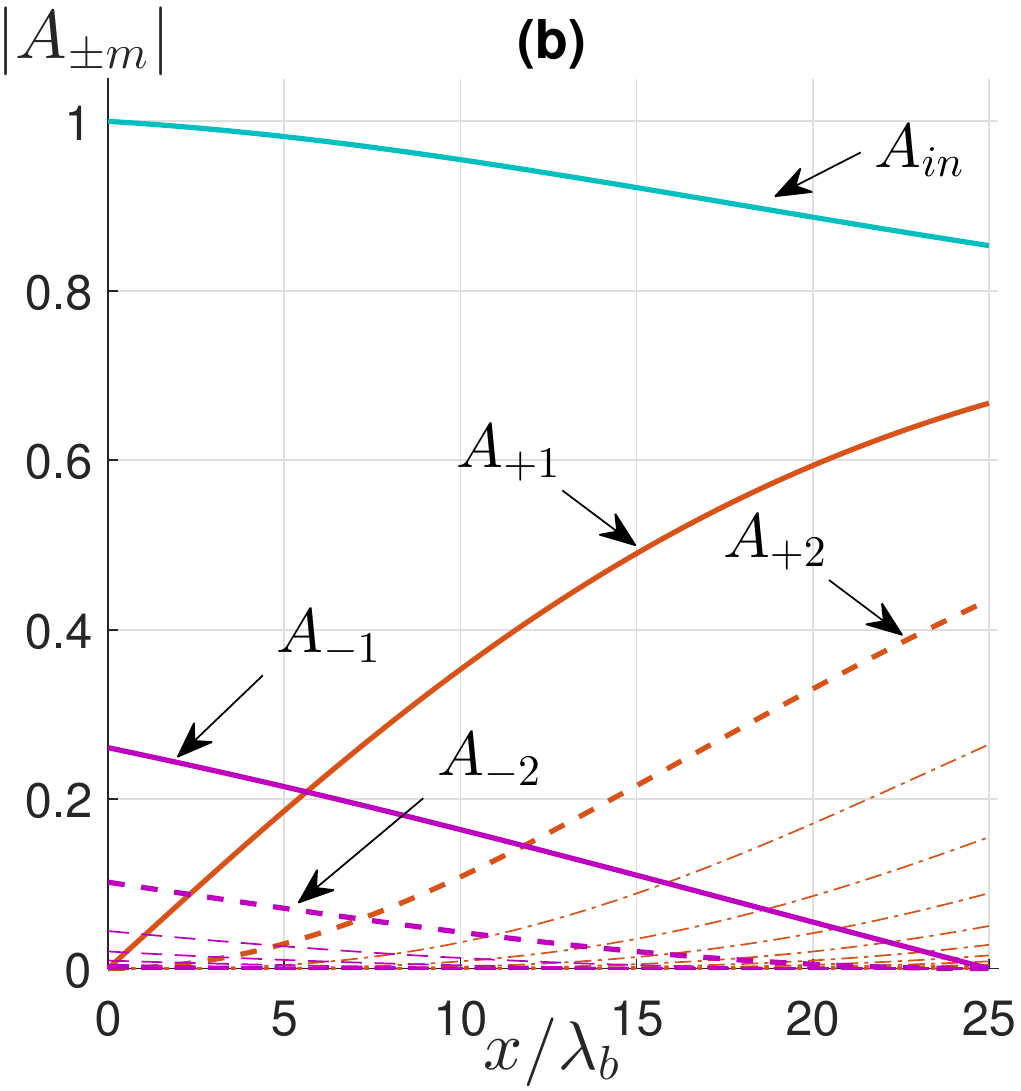}
\hspace{-0.0in}\includegraphics[width=0.33\textwidth]{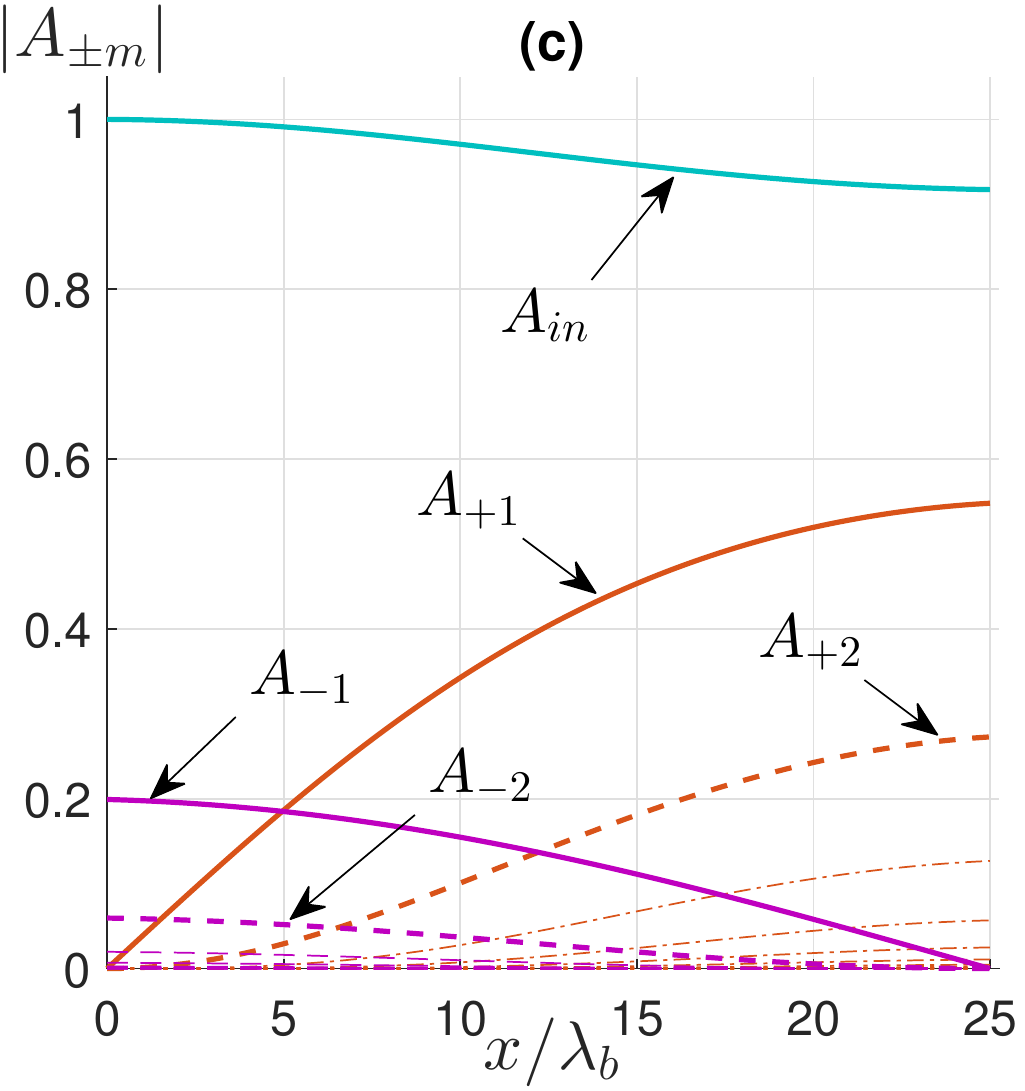}
\vspace{-0.15in}
\caption{Same as figure \ref{detfor}, except that this time we have backward-going resonated ($A_{-m}$) waves in addition to forward-going  ($A_{+m}$) waves as we consider shorter corrugations, i.e. such that $\kab\sim 2\kain$. The first resonated wave $A_{+1}$ is a mode-three wave, and $A_{-1}$ is a mode-one wave. The detuning $\epsilon\delta_{+1}/\kain$ is again (a) 0, (b) 2\%, and (c) 4\%. }\label{detback}
\end{figure}

We show in figure \ref{detback} the chain resonance when the normally-incident waves and the corrugations first resonate a back-scattered mode-one wave $A_{-1}$, propagating in the $-x$ direction, in addition to the forward-going mode-three wave $A_{+1}$ (which is the situation that corresponds to corrugations roughly half the size of those considered in figure \ref{detfor}, i.e. $\kab\sim 2\kain$). The first resonance condition is perfectly satisfied with $\kab=2\kain$, hence so are the successive resonances (cf. figure \ref{detback}a). A cascade of energy is again obtained from the longest to the smallest scales. Back-scattered waves have zero amplitude at the end of the patch, but grow monotonically to the upstream. With detuning, as in figure \ref{detfor}, we see that fewer waves are resonated, but overall the effect of detuning up to 4\% is relatively small such that many new waves can be resonated. Note that the effect of detuning is smaller for the case with back-scattered waves compared to the case of figure \ref{detfor}b-c partly because $\epsilon\delta_{\pm 1}/\kab$ is smaller in figures \ref{detback}b-c due to the larger $\kab$.

\subsection{Chain resonance at oblique incidence}\label{ss2}

An oblique incidence of the mode-one waves results in non-zero detuning wavenumbers for most resonated waves, even when the incident wave is perfectly tuned (see figure \ref{resodiag}b): we call this the angular detuning. To isolate the effect of the angular detuning from the incident detuning (between the incident waves and the corrugations), here we first assume that the first forward-going forced wave is perfectly resonated, i.e. $\epsilon\delta_{+1}=0$. 

\begin{figure}
%\centering
\hspace{-0.05in}
%\includegraphics[width=0.33\textwidth]{tet052det0.eps}
%\hspace{-0.in}\includegraphics[width=0.33\textwidth]{tet079det0.eps}
%\hspace{-0.0in}\includegraphics[width=0.33\textwidth]{tet105det0.eps}
\includegraphics[width=0.33\textwidth]{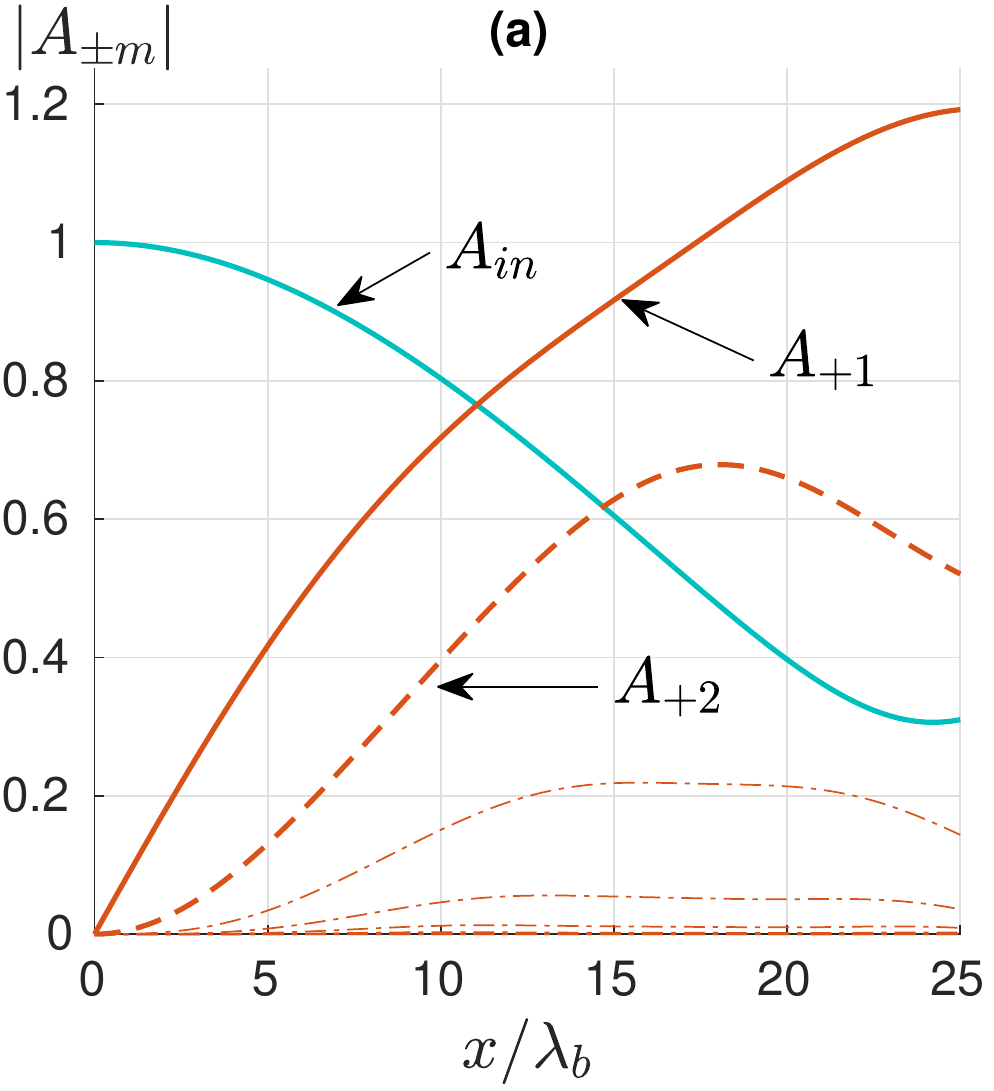}
\includegraphics[width=0.33\textwidth]{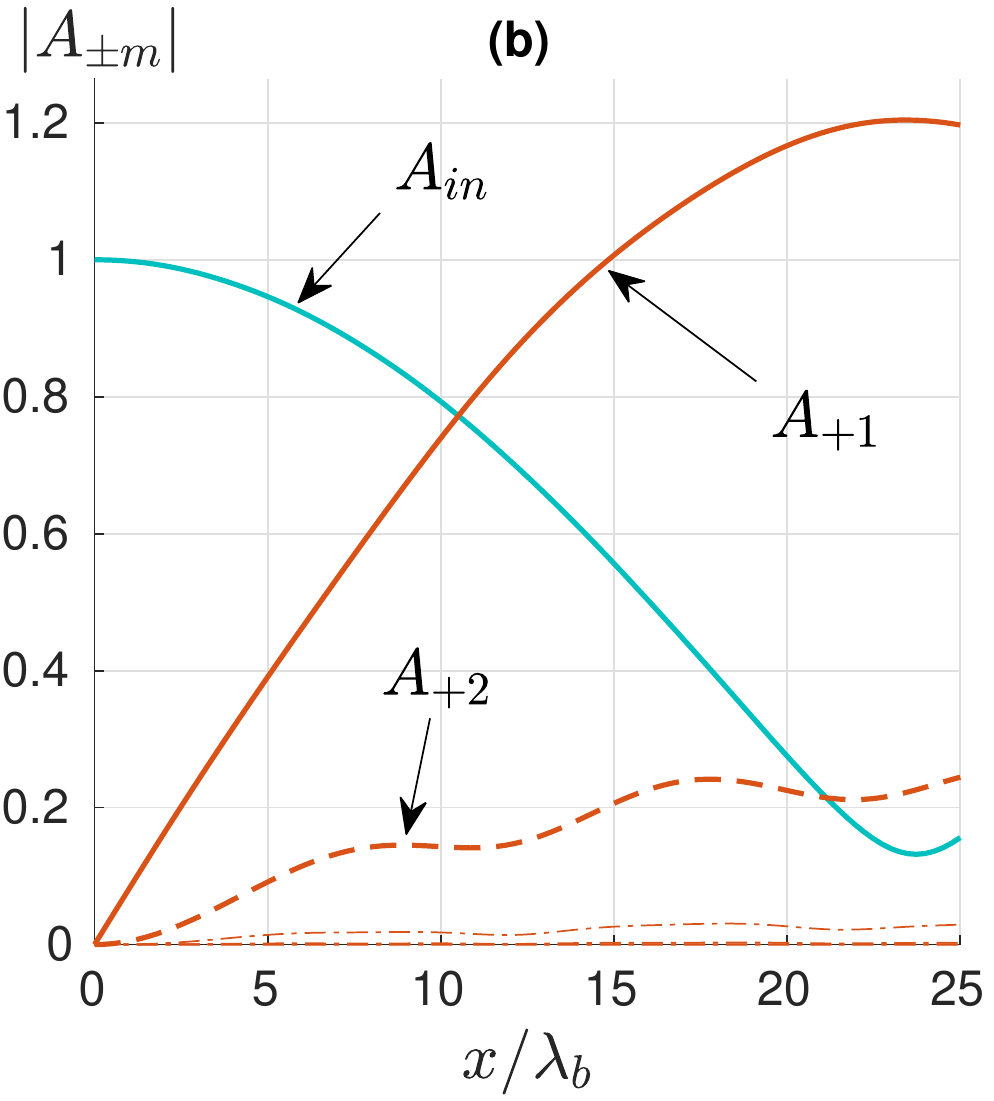}
\includegraphics[width=0.33\textwidth]{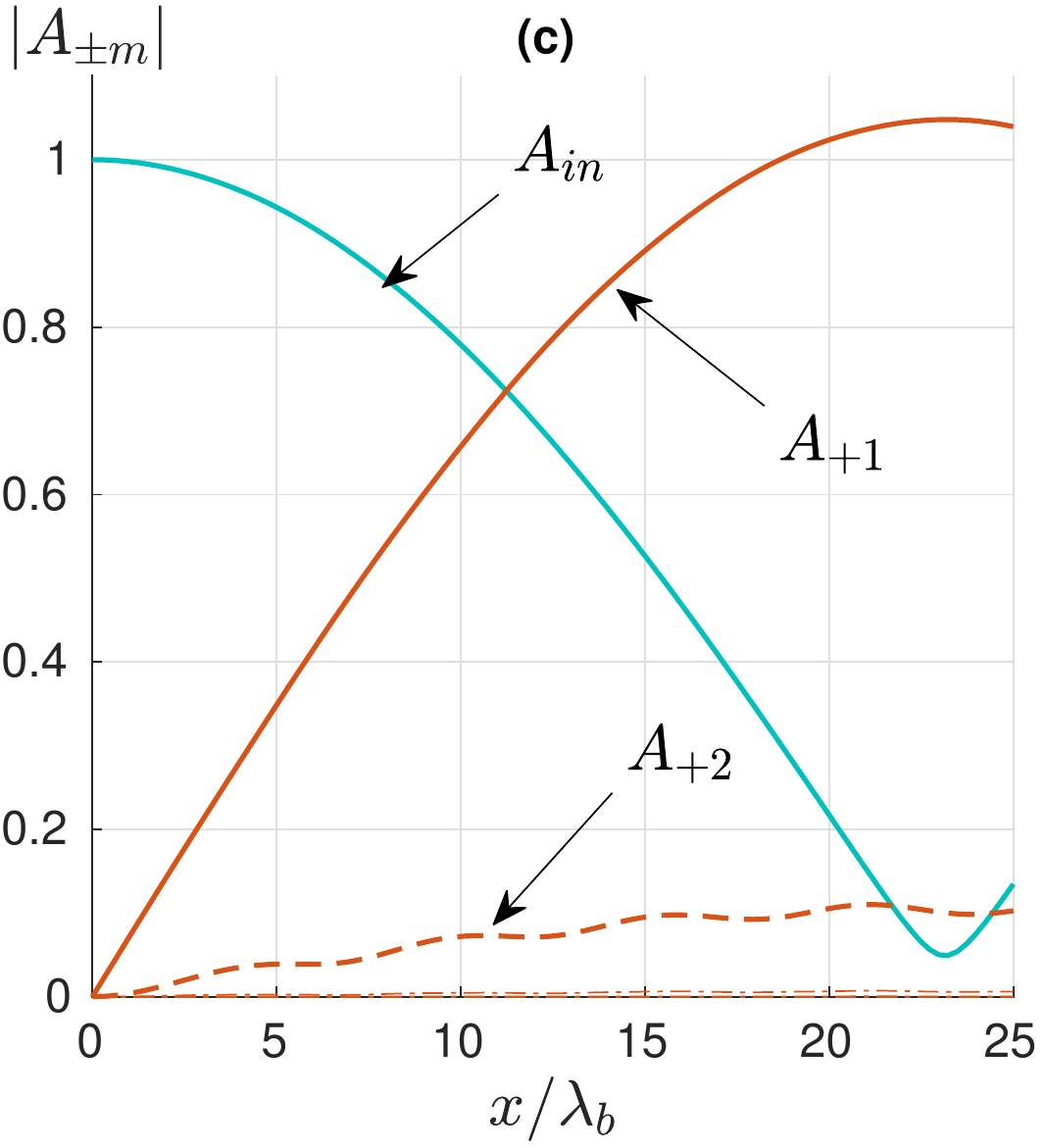}
\vspace{-0.15in}
\caption{Variations of the incident and resonated wave envelopes for incidence angles (a) $\theta_{in}=\pi/6$, (b) $\theta_{in}=\pi/4$, (c) $\theta_{in}=\pi/3$. The first resonance resuling in the mode-two wave $A_{+1}$ is perfect ($\epsilon\delta_{+1}=0$). Back-resonated ($A_{-m}$) waves have negligibe amplitude in all three cases due to large detuning that damp energy exchanges, and therefore are not shown. As in figure \ref{detfor}, $\mathcal{N}=25$, $f/\omega=0.6$, $\kain=\pi$, but the corrugation heigh is this time increased to $d/H=0.015$.}\label{obfor}
\end{figure}

For a relatively broad range of incidence angles, i.e. $\theta_{in}<\pi/8$, we find that the chain resonance results are very similar to the normal-incidence case with no detuning ($\epsilon\delta_{+1}/\kain$). For $\theta_{in}=\pi/6$, figure \ref{obfor}a shows a number of new waves attaining large amplitude over the patch, suggesting that the chain resonance still takes place despite an individual detuning $\epsilon\delta_{+m}/\kain$ and cumulative detuning $\epsilon\Delta_{+m}/\kappa_{r_{+m}}$ increasing up to about 2\% for the first 10 waves. The detuning for the back-scattered waves is quite large (up to 50\% for $A_{-1}$), but even when we consider them, their predicted amplitude is negligible (less than 0.01), such that the results at leading-order do not violate multiple scales theory.

For larger incidence angle $\theta_{in}=\pi/4$, figure \ref{obfor}b shows that the chain resonance is effectively stopped at just 2 or 3 resonated waves. In this case, individual $\epsilon\delta_{+m}/\kain$ and cumulative $\epsilon\Delta_{+m}/\kappa_{r_{+m}}$ detuning wavenumbers go up to 15\% and 4\% for the first 4 resonated waves, and are higher for further resonances. As for the previous case (figure \ref{obfor}a), the detuning of the back-scattered waves is large such that no back-scattered waves are resonated. Figure \ref{obfor}c shows that the two resonated forward-going waves ($A_{+1}$ and $A_{+2}$) are of even smaller amplitudes when the incidence angle is increased to $\theta_{in}=\pi/3$.

\begin{figure}
%\centering
\hspace{-0.05in}
\includegraphics[width=0.33\textwidth]{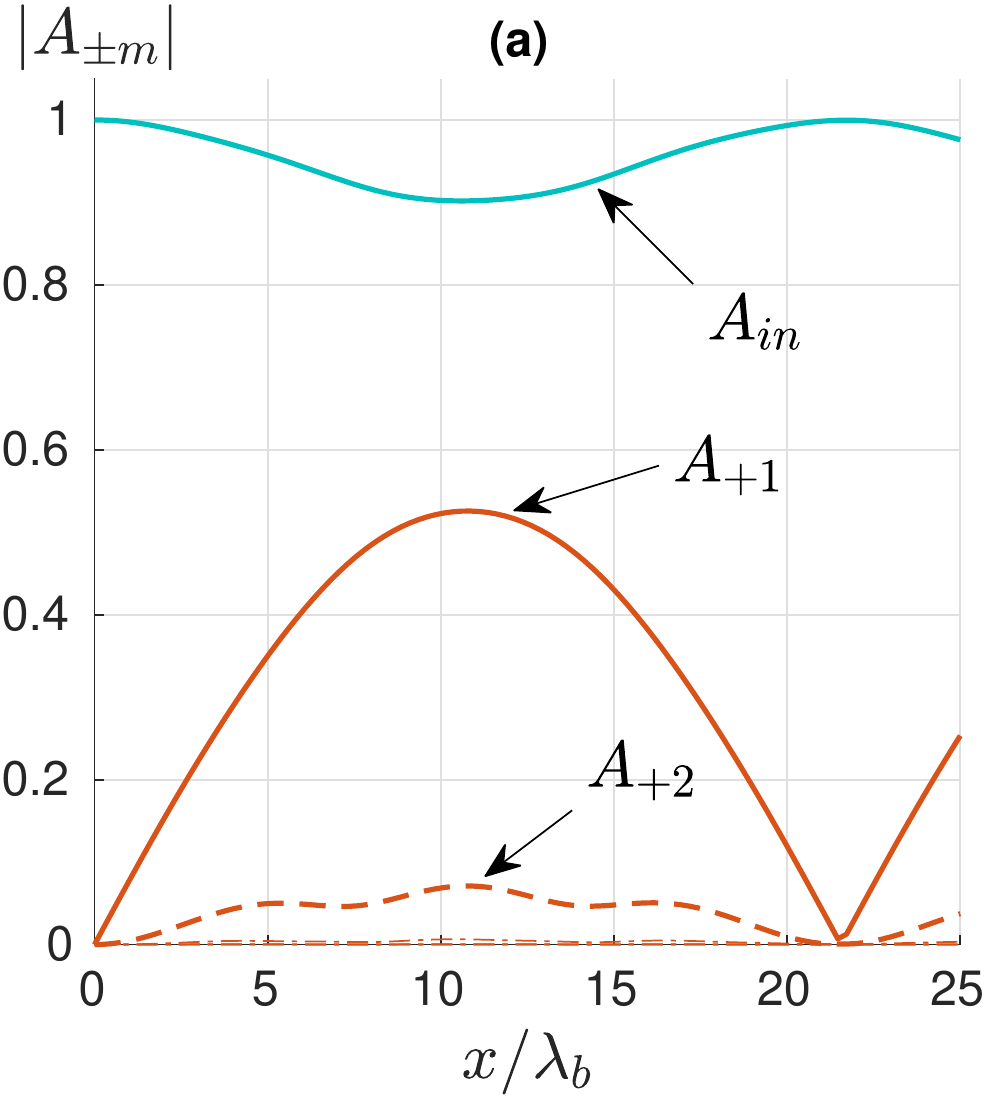}
\hspace{-0.in}\includegraphics[width=0.33\textwidth]{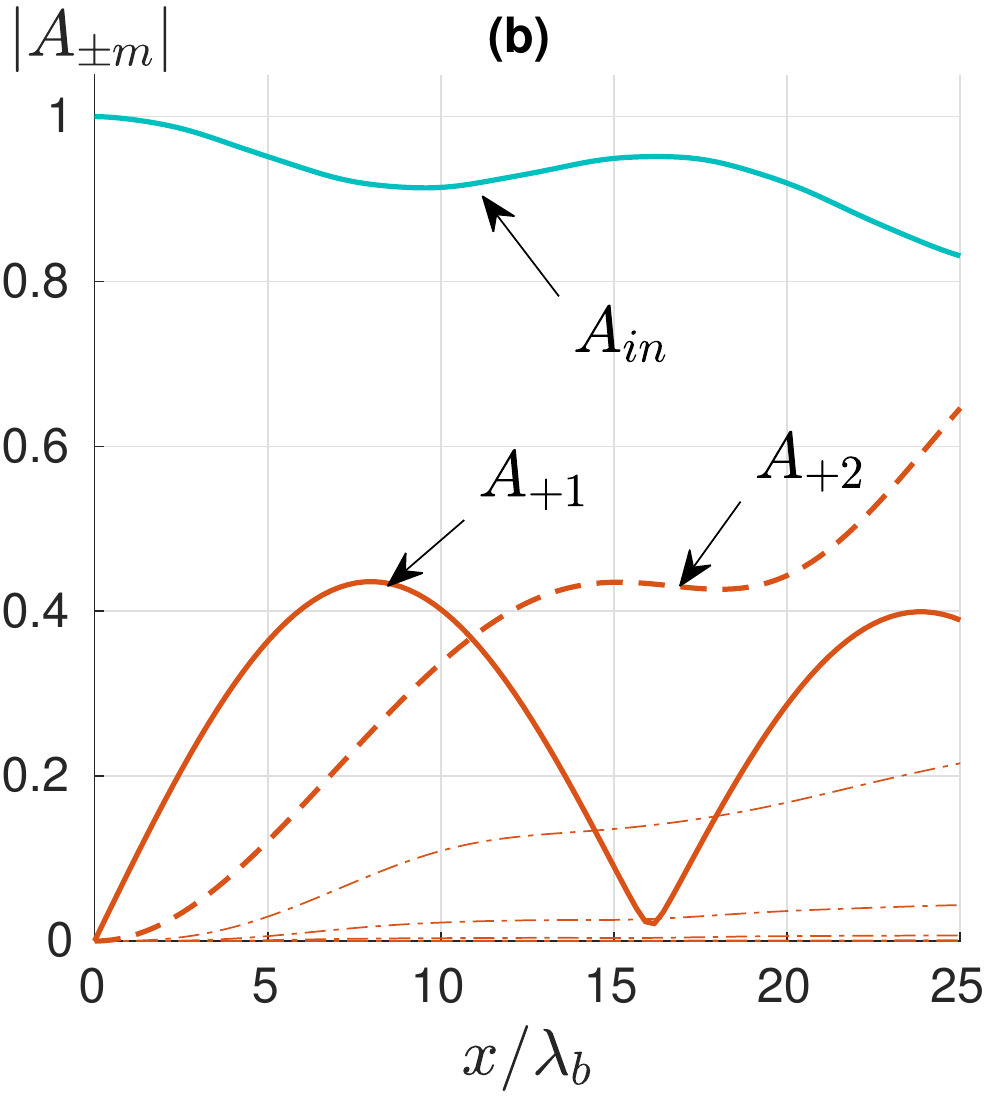}
\hspace{-0.0in}\includegraphics[width=0.33\textwidth]{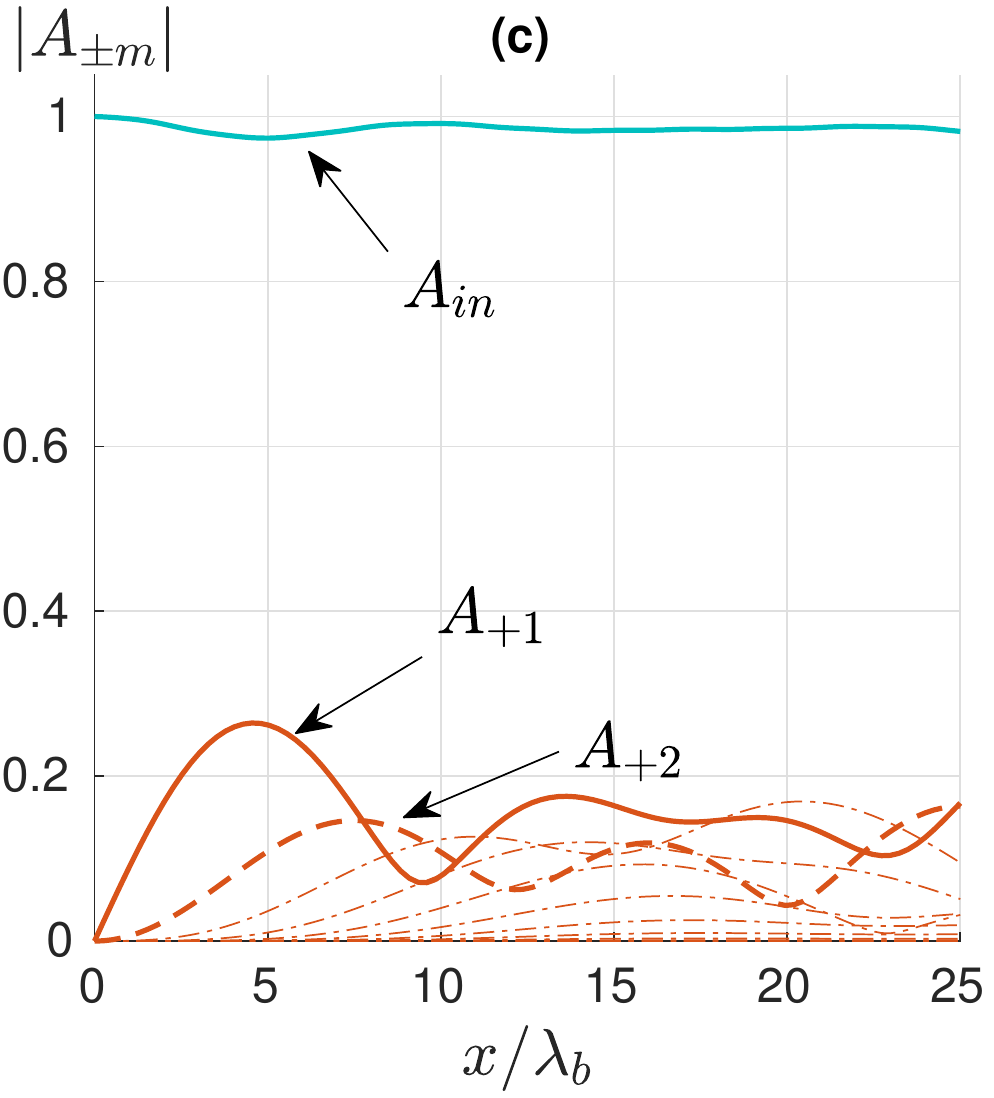}
%\vspace{-0.15in}
\caption{Same case as figure \ref{obfor}b, except that this time the incident-wave detuning $\epsilon\delta_{+1}/\kain$ is 5\% for (a) and (b), and 10\% for (c). Unlike in other cases, in (a) the detuning is negative, i.e. $\epsilon\bm{\delta_{+1}}\cdot \bm{\kappa}_{r_{+1}}<0$ [cf. equation \eqref{bi9}]. In (a) the effects of the incident detuning and the angular detuning are added, such that fewer waves of smaller amplitudes are obtained compared to the perfectly-tuned incident-wave case of figure \ref{obfor}b. In (b) and (c), the two effects are opposite such that more waves are obtained resulting in an oblique chain resonance of waves with different angles of propagation. }\label{obdet}
\end{figure}

In figure \ref{obdet}a-c we show the same case of figure \ref{obfor}b, i.e. with incidence angle $\theta_{in}=\pi/4$, but this time we consider a detuned incident wave. In figure \ref{obdet}a we show the results obtained for $\epsilon\bm{\delta}_{+1}=-0.025\bm{\kappa}_{r_{+1}}$ (i.e. $\epsilon\delta_{+1}=0.05\kain$), corresponding to incident waves that are too short to perfectly resonate the $\bm{\kappa}_{r_{+1}}$ waves. Fewer waves are resonated and have smaller maximum amplitude than in figure \ref{obfor}b. The effect of the incident-wave detuning can thus be seen to be added to that due to the oblique incidence. The oblique angular detuning is indeed  opposite to the direction of the $\bm{\kappa}_{r_{+m}}$ waves  (as is the case of the incident detuning), which makes the wave-bottom interactions always result in a wavenumber larger than the next closest natural resonated wavenumber $\bm{\kappa}_{r_{+(m+1)}}$. When the incident-wave detuning is $\epsilon\bm{\delta}_{+1}=0.025\bm{\kappa}_{r_{+1}}$ (cf. figure \ref{obdet}b), however, it counteracts the angular detuning such that more waves with relatively large amplitudes are resonated and a chain resonance is again obtained. With a larger positive detuning of $\epsilon\bm{\delta}_{+1}=0.05\bm{\kappa}_{r_{+1}}$, even more waves are resonated but the maximum amplitudes  of the low-mode waves over the patch are smaller (figure \ref{obdet}c). The trend continues as the detuning is increased, until no resonances remain.

\subsection{Discussion of results}\label{ss3}

\begin{figure}
\centering
%\hspace{-0.285in}%
\includegraphics[width=1.\textwidth]{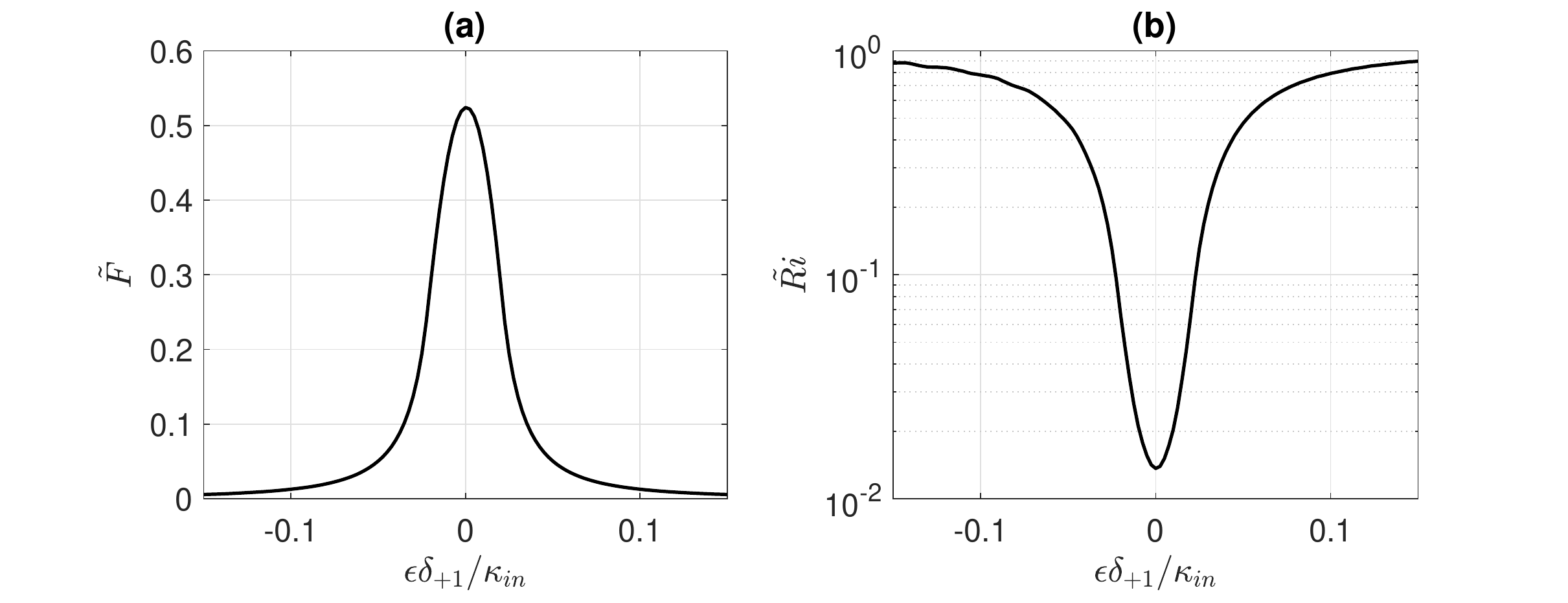}
\vspace{-0.15in}
\caption{ Effect of detuning on (a) the maximum of energy flux of the resonated waves $\tilde{F}$ (equation \eqref{dec}), and (b) the minimum Richardson number $\tilde{Ri}$ (see equation \eqref{ri}). The results are shown as a function of the normalized detuning wavenumber $\epsilon\delta_{+1}/\kain$ for the case of normal incidence ($\theta_{in}=0$). As in figure \ref{detfor}, $\kain=\pi$, $\mathcal{N}=25$, $f/\omega=0.6$, and $d/H=0.008$.  }\label{Rich0}
\end{figure}

From the various results shown in figures \ref{detfor}-\ref{obdet} it is found that the resonant generation of multiple internal waves over seabed corrugations is not limited to the special case of perfectly-tuned normally-incident waves, although the efficiency of energy transfers does change with changing incidence angle and detuning.. The sensitivity of the resonance to the incident detuning can be partly assessed from the variations with $\epsilon\delta_{+1}/\kappa_{in}$ of
\ba{}\label{dec}
\tilde{F}=\max_x\lp 1-|A_{in}|^2\rp,
\ea
which is the maximum energy flux of the resonated waves in $x$ direction  over the patch (normalized by the incident energy flux; see equation \eqref{flux}). For normal incidence ($\theta_{in}=0$), figure \ref{Rich0}a shows that the resonated waves capture more than 50\% of the incident energy flux with zero detuning and more than 40\% for a broad range of detuning wavenumber. The peak as well as the half width of this curve would be increased provided that the corrugation amplitude increases too; the peak would also increase with the number of corrugations. While figure \ref{Rich0}a implies that strong energy exchanges occur with or without detuning, it does not yet allow us to assess whether high-wavenumber waves are generated for relatively large detuning. In fact, as shown in figure \ref{obfor}c, the incident wave can cede a significant fraction of its energy to just two resonated waves.

Because generating high-wavenumber waves is expected to increase the vertical shear, here we make a parallel between the efficiency of chain resonance at generating high-wavenumber waves and the decrease of the Richardson number $Ri=(HN/U)^2/(\p V/\p z)^2$ over the patch, where $V$ is the dimensionless horizontal velocity and $z$ the dimensionless vertical coordinate. The Richardson number is of great interest to the oceanographic community as it is often used as a proxy to predict the likelihood of overturning events and enhanced mixing in a stratified fluid, due to shear instability acting against the stable buoyancy field. Though the existence of a universal critical Richardson number $Ri_c$ at which turbulent mixing can start is still debated, the lower bound $Ri<Ri_c=1/4$, originally derived for horizontal shear flows, or the order of unity is often assumed \cite[][]{Mack2004,Galperin2007}. The Richardson number for mode-one internal waves at tidal frequency can be roughly estimated from field measurements in the open ocean. Velocities of mode-one semidiurnal internal waves have been reported to be of order $U\sim\co(0.01-0.1 ms^{-1})$ and higher close to large topographic features \cite[][]{Garrett2007,Alford2007a,Alford2007}, such that for a depth of $H=4000m$ and constant buoyancy of $N=2\times 10^{-4} s^{-1}$  \cite[which is within the range reported in some parts of the deep ocean, see e.g.][]{StLaurent2002}, the minimum Richardson number is of order $Ri=(NH/U)^2/\pi^2 \sim \co(10)-\co(10^3) \gg 1$. Low-mode internal tides at the semidiurnal frequency are thus unlikely to overtun because of shear instability, which explains why it is generally assumed that a nonlinear mechanism or modal transformation must take place in order to destabilize the flow and induce mixing \cite[][]{Staquet2002}.

Here with simple monochromatic corrugations we showed that an efficient modal transformation can occur for a wide range of incidence angles and detuning. The efficiency of the energy transfer from low-mode to high-mode internal waves is now related to the drop of Richardson number $\tilde{Ri}$, i.e. 
\ba{}\label{ri}
\tilde{Ri}=   \f{\lp 1+\f{f^2}{\omega^2}\rp \kain^2  }{\max_{x,z}\lb\p_z\sqrt{|u^{(0)}|^2+|v^{(0)}|^2}\rb^2} ,
\ea
which gives the minimum of the Richardson number over the corrugations normalized by the Richardson number of the incident wave. For the case of normal incidence (see figure \ref{Rich0}b), we find that $\tilde{Ri}$ is close to $10^{-2}$, with no detuning. This is a relatively large decrease of the Richardson number. However, whether it is a sufficient decrease for the waves to overturn depends on the Richardson number of the initial wave (say $Ri_0$), since the actual Richardson number is $Ri_0\times\tilde{Ri}$.  Interestingly, the decrease of Richardson number can be used to obtain a lower bound on the wavenumber of the short internal waves generated by the chain resonance. Assuming that all of the incident energy flux is sent to the $\kappa_{r_{+m}}$ wave only, the wave amplitude becomes $\kappa_{r_{+m}}^{1/2}$ such that $\tilde{Ri}\sim (2\kain)^2/(4\kappa_{r_{+m}}^{5/2})$, or $\tilde{Ri}\sim 1/(m+1)^{5/2}$ if $\kappa_{r_{+m}}=(m+1)\kain$ (as is the case for figure \ref{Rich0}). Therefore, a drop of $10^{-2}$ implies the lower bound $m=5$ for the number of resonated waves; it can be checked from figure \ref{detfor}a that at least 5 waves are resonated with non-negligible amplitude. %Again, the effect of detuning is relatively small and the lowest Richardson number is obtained for a relativel broad range of detuning wavenumber. 

\begin{figure}
\centering
%\hspace{-0.285in}%
\includegraphics[width=0.95\textwidth]{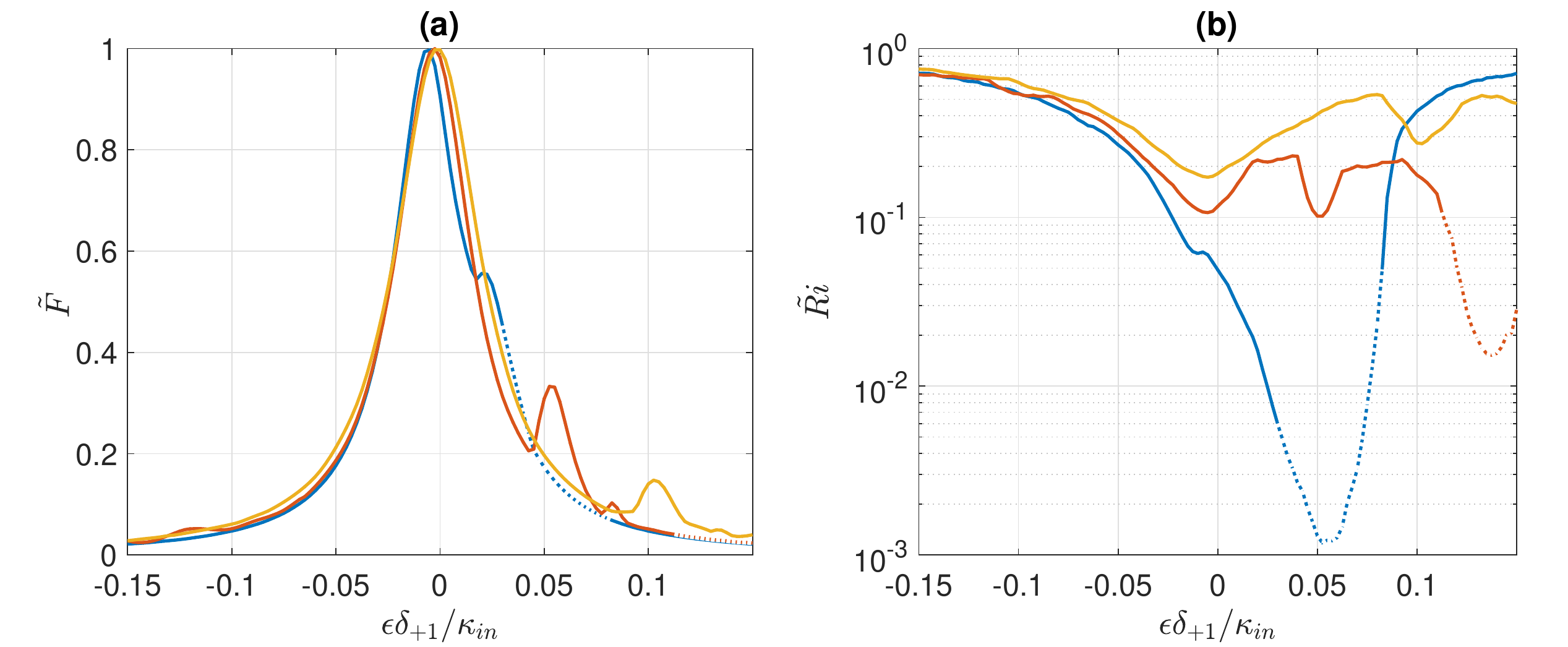}
\vspace{-0.05in}
\caption{Same as figure \ref{Rich0} but for incidence angle $\theta_{in}=\pi/6$ (\blue\L), $\pi/4$ (\red\L), $\pi/3$ (\yellow\L), and $d/H=0.015$. The energy flux ceded to the resonated waves is close to unity in all three cases when $\delta_{+1}\sim 0$. $\tilde{Ri}$ decreases significantly for $\theta_{in}=\pi/6,\pi/4$, although the location of the global minimum is obtained for non-zero detuning; this implies that the effect of oblique incidence and incident detuning can cancel each other, resulting in the generation of multiple higher-wavenumber resonated modes. Parts of the curves shown as dotted lines correspond to cases for which the theory is relatively stretched, and therefore may not be quantitatively accurate.}\label{Rich}
\end{figure}

We show in figure \ref{Rich}a the maximum energy flux of the resonated waves and in figure \ref{Rich}b the minimum Richardson number for oblique incident waves ($\theta_{in}=\pi/6,\pi/4,\pi/3$). The maximum energy flux of the resonated waves is large (even higher than in the case of normal incidence, see figure \ref{Rich0}a), partly because the corrugation amplitude is larger, but also because the energy transfer from the incident to the resonated waves can occur more rapidly with only one wave resonated. In figure \ref{Rich}b we observe that there exists an incident detuning for which the minimum Richardson number drops by several orders of magnitude for $\theta_{in}=\pi/6,\pi/4$, as was the case for normal incidence. Again, the occurrence of chain resonance for non-zero detuning at oblique incidence comes from the fact that the angular detuning is negative, hence can be counter-balanced with a positive detuning of the first resonance. Because the first interaction that feeds all subsequent resonances is weakened with non-zero incident detuning, the resonated waves  have relatively smaller amplitudes in chain resonance at oblique incidence (for which we require a non-zero incident detuning) than in chain resonance at normal incidence (obtained with zero incident detuning). At oblique incidence, however, the projected patch length onto the internal waves' path is longer than at normal incidence, hence resulting in the generation of more waves (this is the effect of the $\cos\tm$ terms in \eqref{bi19}). These two effects compete with each other, allowing for significant decrease of the Richardson number at oblique incidence (cf. figure \ref{Rich}b). The fact that the minimum Richardson number is obtained for increasing detuning as the incidence angle is increased is simply the effect of the increased angular detuning that must be counter-balanced.

\subsection{Extention to polychromatic seabed}\label{ss4}

In this section we briefly discuss the extension of the present theory to the case of a polychromatic seabed, i.e. with corrugations given by (in dimensionless form)
\ba{}\label{ex1}
b(x)=\sum_{n=1}^{N_c} \f{d_n}{2H} e^{i\varphi_n}e^{i\kappa_{bn}x}+c.c.
\ea
In \eqref{ex1} the bottom wavenumbers ${\kappa_{bn}}$ are ordered from smallest to largest, $d_n$ is real ($d_n/H \ll 1$) and $\varphi_n$ is the mode phase. For simplicity we restrict the analysis to one horizontal dimension, no rotation, and we consider the discretized bathymetric spectrum to be narrow-banded, such that all seabed modes resonate the same sequence of internal waves $\{\kappa_{r_{\pm m}}\}$ for a given incident wavenumber $\kain$. In other words we assume that all wavenumbers in $[\kappa_{b1},\kappa_{bN}]$ are closest to the same perfectly-resonant bottom wavenumber $\kappa_b^*$.%, and so the next closest perfectly-resonant bottom wavenumber must lay outside $[2k_{b1}-k_b^*,2k_{bN}-k_b^*]$. 

Equation \eqref{ex1} can be recast advantageously as
\ba{}\label{ex1b}
b(x)=e^{i\kappa_b^*x}\sum_{n=1}^{N_c} \f{d_n}{2H} e^{i\varphi_n}e^{-i\alpha_{n}x}+c.c.
\ea
with $\alpha_n $ a bottom detuning representing the deviation of each bottom wavenumber from $\kappa_b^*$. For a narrow-banded topographic spectrum we have $|\alpha_n| \ll \kappa_b^* \sim 1$, such that the sum term in \eqref{ex1b} is a slowly-varying function of $x$. Therefore, considering a narrow-banded polychromatic seabed is the same as considering a slowly-varying and oscillating bottom corrugation amplitude $d\equiv D(x)$, such that the envelope equation \eqref{bi18} can be readily modified as
\ba{}\label{ex6}
\f{\d A_{\pm m}}{\d x}  = i \kam \lb  A_{\pm(m+1)} \I_{\pm}(x)  + A_{\pm(m-1)} \I_{\mp}(x)  \rb
\ea
with the slowly-varying interaction term 
\ba{}\label{ex7}
\I_{\pm}(x)  = \sum_{n=1}^{N_c} \f{d_n}{2H} e^{\pm i\varphi_n} e^{\mp i\alpha_n x},
\ea
for bottom corrugations of the form \eqref{ex1}. It should be noted that equations \eqref{ex6}-\eqref{ex7} can be derived formally as in \S\ref{sec:bi3} provided that a superposition of $N_c$ interactions is considered on the right-hand-side of the bottom boundary condition \eqref{bi133}; while necessary for broad-band spectra, this procedure is, however, not justified for the present case.

\begin{figure}
\centering
%\hspace{-0.285in}%
\includegraphics[width=0.8\textwidth]{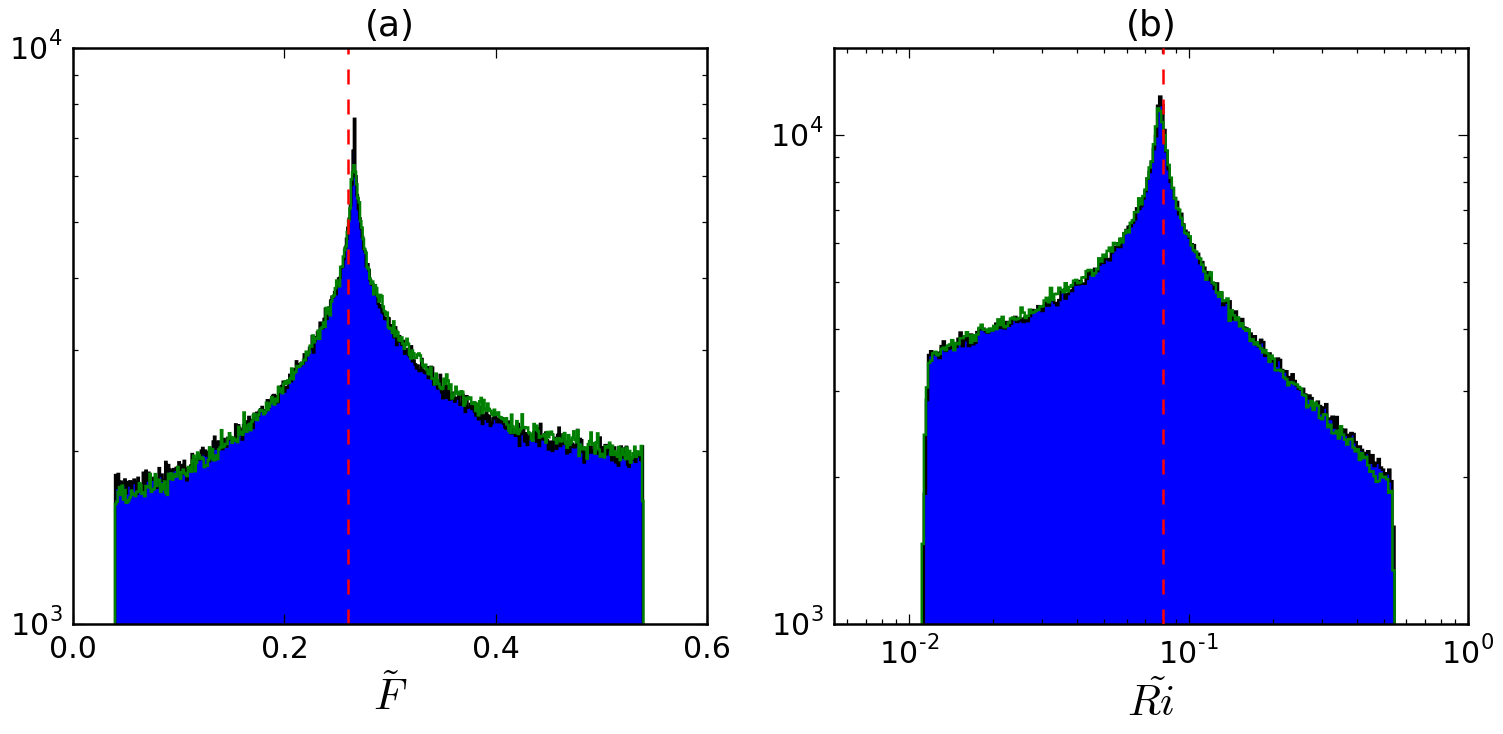}
\vspace{-0.05in}
\caption{Histograms of (a) $\tilde{F}$ (equation \eqref{dec}), and (b) $\tilde{Ri}$ (equation \eqref{ri}) obtained for $10^6$ distinct uniform random bottom phase distributions $\varphi_n\in[0,2\pi[$ (see equation \eqref{ex1}). 3 bottom modes are considered with amplitudes $d_n/H=[0.0025,~0.005,~0.0025]$, and detuning $\alpha_n/\kain=[-0.02,~0,~0.02]$. The incident wave is normal to the $\mathcal{N}=25$ corrugations considered. The vertical dashed  lines show the solutions when the two detuned bottom modes are discarded, and the green solid line shows the histogram obtained with 2 added bottom modes of amplitudes $0.0025$ and detuning $\pm 0.08$. $f/\omega=0$. }\label{distributions}
\end{figure}

Equations \eqref{ex6}-\eqref{ex7} demonstrate that the resonance mechanism due to multiple seabed components can be described as a superposition of resonances (from the sum term). The interference effects resulting from the superposition can be destructive or constructive depending on the bottom phases $\varphi_n$, detuning $\alpha_n$, and location $x$ over the corrugations. Some results on the sensitivity of the chain resonance to arbitrary bottom-phase distributions are provided in figure \ref{distributions}. We consider 3 bottom modes and solve equation \eqref{ex6} for $10^6$ different cases of uniform random phase distributions $\varphi_n\in[0,2\pi[$. Figures \ref{distributions}a,b show the histograms of the maximum energy flux ceded by the incident wave to the resonated waves ($\tilde{F}$), and of the minimum normalized Richardson number ($\tilde{Ri}$) for the $10^6$ random phase distributions (cf. equations \eqref{dec}-\eqref{ri}). The most probable values of the histograms of $\tilde{F}$ and $\tilde{Ri}$ are approximately equal to what is obtained when only the perfectly-tuned bottom mode is considered (shown by the dashed vertical lines). The histograms are skewed toward higher/smaller values of $\tilde{F}$/$\tilde{Ri}$, such that adding bottom modes with random phases is found to lead to constructive interference (higher $\tilde{F}$, lower $\tilde{Ri}$) with slightly higher probabilities than destructive interference. This suggests that the chain resonance mechanism can be expected to occur in most cases of topographic spectra even though the intensity is as expected sensitive to the phase distribution. The minimum/maximum values of $\tilde{F}$ and maximum/minimum values of $\tilde{Ri}$ are obtained when the interference lead to the minimum/maximum possible  transfer of energy to the shorter waves. With two bottom modes of relatively large detuning added (light green solid lines), it is seen that the histograms do not change significantly. This is expected since the higher the detuning the lesser the contribution to the generation of new resonated waves.

\section{Concluding remarks}\label{sec:bi5}

In this work we derived the envelope equations \eqref{bi19}-\eqref{bi19b} describing energy transfers from an incident low-mode internal wave to shorter waves in space over resonant seabed topography. We considered the case of a linearly-stratified rotating Boussinesq fluid with a top rigid lid, and we focused on monochromatic small-amplitude seabed corrugations as the resonant bathymetry.  Interference effects obtained when considering a polychromatic seabed with arbitrary phase distributions were briefly discussed in \S\ref{ss4}.  

For a broad range of incidence angles of the mode-one internal wave, we demonstrated that a large number of other freely propagating waves become generated as a result of wave-wave resonances occurring in series. The successively resonated waves are of increasing wavenumber, and, for oblique incidence of the incident mode-one wave, they propagate in many different directions. The resonated waves naturally experience some detuning when (i) the incident wave is oblique with respect to the corrugations (because then the resonant wavenumbers are not uniformly spaced, this is called angular detuning), and (ii) when the first interaction between the incident mode-one wave and the corrugations is detuned (which we called incident detuning). The spatial variations of the envelope amplitudes are a priori arbitrary, but turn out to be periodic when the number of resonated waves is finite (note, however, that the periodicity is only clearly seen for a long-enough corrugated patch, cf. e.g. figure \ref{obdet}a).

Because near-resonance waves are as important as perfectly-resonant waves, the effect of the incident detuning must be thoroughly considered. Unlike the classical Bragg resonance of surface waves in the homogeneous ocean, we showed that the effect of the incident detuning can be complex and potentially lead to enhanced wave generation when it balances off the angular detuning at oblique incidence. 

An important limitation of the model comes from the asymptotic expansion of the bottom boundary condition, which requires large internal wavelengths in the vertical direction compared to the corrugation amplitude. This implies that one must consider corrugations with quite small amplitudes, and therefore in large number, to obtain and correctly predict the envelope variations of high-wavenumber waves in a chain resonance. Fully linear models that do not rely on an asymptotic expansion of the bottom boundary condition are not subject to this limitation, but become significantly computationally expensive in three dimensions. This is why the present leading-order multiple scales model is expected to advance our understanding of resonant scattering events in the oceans for many different physical conditions.  

We expect the extension of our results to the case of arbitrary stratification to be relatively straightforward. The number of resonated waves as well as the intensity of energy exchanges in the chain resonances will differ from one stratification profile to another, but the solution method stays the same. The effect of a free-surface, which we investigated but did not report here, has in general very little effect on the chain resonance obtained. The wave mode solutions of the linear problem are the same as with a rigid-lid, and only the wavenumbers become slightly shifted. The possibility to resonate back-scattered waves via the surface-wave mode for long corrugations (i.e. $k_b\sim k_{in}$), however, is unique to stratified fluids with a free top boundary.

Even though most of our results were presented for monochromatic long-crested corrugations ($\Kab=\kab\hat{x}$), the theory can be easily extended to multiple corrugation wavenumbers with arbitrary orientations in $(x,y)$ space, in which case one can expect a ramification of the chain resonance across the two-dimensional wavenumber space. The number of resonant branches may in fact become infinite with just two seabed harmonics and may exhibit interesting different chain resonance dynamics. In order to extend the range of applicability of the theory to more general cases, finite-amplitude bottom corrugations should be considered and are worth independent investigation. However, large seabed variations require higher-order or exact linear wave theories, such as Green functions approach or Floquet theory, making the problem much less tractable analytically \cite[see e.g. in homogeneous water][]{Yu2012a}.

The redistribution of energy across the internal wave spectrum near topography is of significant importance when it leads to the occurrence of enhanced mixing events and wave breaking. Therefore, it will be  essential to include the effect of finite-amplitude internal waves in future studies. Laboratory experiments and numerical simulations will be particularly helpful in extending the present predictions to fully nonlinear wave regimes, and in considering relatively large seabed corrugations over which the chain resonance of internal waves should occur rapidly.  

\section{Acknowledgment}

The authors would like to thank the support from the American Bureau of Shipping. L.-A. Couston was supported by a Frank and Margaret Lucas Scholarship while at the University of California, Berkeley. We would also like to thank the anonymous reviewers for their constructive suggestions and comments, which greatly improved the paper.

%%%%%%%%%%%%%%%%%%%%%%%%%%%%%%%
%%%%%%%%%%%%%%%%%%%%%%%%%%%%%%%
\bibliographystyle{jfm}
\bibliography{MainBiblio}

\begin{thebibliography}{51}
\expandafter\ifx\csname natexlab\endcsname\relax\def\natexlab#1{#1}\fi

\bibitem[Alam(2012)]{Alam2012}
{\sc Alam, M.-R.} 2012 {A new triad resonance between co-propagating surface
  and interfacial waves}. {\em J. Fluid Mech.\/} {\bf 691}, 267--278.

\bibitem[Alam {\em et~al.\/}(2009)Alam, Liu \& Yue]{Alam2009}
{\sc Alam, M.-R., Liu, Y. \& Yue, D. K.~P.} 2009 {Bragg resonance of waves in a
  two-layer fluid propagating over bottom ripples. Part I. Perturbation
  analysis}. {\em J. Fluid Mech.\/} {\bf 624}, 191--224.

\bibitem[Alam \& Mei(2007)]{Alam2007a}
{\sc Alam, M.-R. \& Mei, C.~C.} 2007 {Attenuation of long interfacial waves
  over a randomly rough seabed}. {\em Journal of Fluid Mechanics\/} {\bf 587},
  73--96.

\bibitem[Alford {\em et~al.\/}(2007)Alford, MacKinnon, Zhao, Pinkel, Klymak \&
  Peacock]{Alford2007}
{\sc Alford, M.~H., MacKinnon, J.~A., Zhao, Z., Pinkel, R., Klymak, J. \&
  Peacock, T.} 2007 {Internal waves across the Pacific}. {\em Geophysical
  Research Letters\/} {\bf 34}~(24), 2--7.

\bibitem[Alford \& Zhao(2007)]{Alford2007a}
{\sc Alford, M.~H. \& Zhao, Z.} 2007 {Global Patterns of Low-Mode Internal-Wave
  Propagation. Part I: Energy and Energy Flux}. {\em Journal of Physical
  Oceanography\/} {\bf 37}, 1829--1848.

\bibitem[Balmforth {\em et~al.\/}(2002)Balmforth, Ierley \&
  Young]{Balmforth2002}
{\sc Balmforth, N.~J., Ierley, G.~R. \& Young, W.~R.} 2002 {Tidal Conversion by
  Subcritical Topography}. {\em Journal of Physical Oceanography\/} {\bf
  32}~(10), 2900--2914.

\bibitem[Becker {\em et~al.\/}(2009)Becker, Sandwell, Smith, Braud, Binder,
  Depner, Fabre, Factor, Ingalls, Kim, Ladner, Marks, Nelson, Pharaoh, Trimmer,
  {Von Rosenberg}, Wallace \& Weatherall]{Becker2009}
{\sc Becker, J.~J., Sandwell, D.~T., Smith, W. H.~F., Braud, J., Binder, B.,
  Depner, J., Fabre, D., Factor, J., Ingalls, S., Kim, S-H., Ladner, R., Marks,
  K., Nelson, S., Pharaoh, A., Trimmer, R., {Von Rosenberg}, J., Wallace, G. \&
  Weatherall, P.} 2009 {Global Bathymetry and Elevation Data at 30 Arc Seconds
  Resolution: SRTM30{\_}PLUS}. {\em Marine Geodesy\/} {\bf 32}~(4), 355--371.

\bibitem[Bell(1975{\natexlab{{\em a\/}}})]{Bell1975b}
{\sc Bell, T.~H.} 1975{\natexlab{{\em a\/}}} {Lee waves in stratified flows
  with simple harmonic time dependence}. {\em Journal of Fluid Mechanics\/}
  {\bf 67}~(4), 705--722.

\bibitem[Bell(1975{\natexlab{{\em b\/}}})]{Bell1975}
{\sc Bell, T.~H.} 1975{\natexlab{{\em b\/}}} {Topographically generated
  internal waves in the open ocean}. {\em Journal of Geophysical Research\/}
  {\bf 80}~(3), 320--327.

\bibitem[B{\"{u}}hler \& Holmes-Cerfon(2011)]{Buhler2011}
{\sc B{\"{u}}hler, O. \& Holmes-Cerfon, M.} 2011 Decay of an internal tide due
  to random topography in the ocean. {\em Journal of Fluid Mechanics\/} {\bf
  678}, 271--293.

\bibitem[B{\"{u}}hler \& Muller(2007)]{Buhler2007}
{\sc B{\"{u}}hler, O. \& Muller, C.~J.} 2007 {Instability and focusing of
  internal tides in the deep ocean}. {\em Journal of Fluid Mechanics\/} {\bf
  588}, 1--28.

\bibitem[Chen(2009)]{Chen2009}
{\sc Chen, E.} 2009 {Degradation of the internal tide over long bumpy
  topography}. {\em Woods Hole GFDL reports\/} pp. 248--268.

\bibitem[Couston {\em et~al.\/}(2017)Couston, Jalali \& Alam]{Couston2017}
{\sc Couston, L.-A., Jalali, M.~A. \& Alam, M.-R.} 2017 {Shore protection by
  oblique seabed bars}. {\em Journal of Fluid Mechanics\/} {\bf 815}, 481--510.

\bibitem[Elandt {\em et~al.\/}(2014)Elandt, Shakeri \& Alam]{Elandt2014}
{\sc Elandt, R.~B., Shakeri, M. \& Alam, M.-R.} 2014 {Surface gravity-wave
  lensing}. {\em Phys. Rev. E Stat. Nonlin. Soft Matter Phys.\/} {\bf 89},
  1--6.

\bibitem[Exarchou {\em et~al.\/}(2012)Exarchou, {Von Storch} \&
  Jungclaus]{Exarchou2012}
{\sc Exarchou, E., {Von Storch}, J.~S. \& Jungclaus, J.~H.} 2012 {Impact of
  tidal mixing with different scales of bottom roughness on the general
  circulation}. {\em Ocean Dynamics\/} {\bf 62}~(10-12), 1545--1563.

\bibitem[Fredholm(1903)]{Fredholm1903}
{\sc Fredholm, I.} 1903 {Sur une classe d'{\'e}quations fonctionnelles}. {\em
  Acta Mathematica\/} {\bf 27}~(1), 365--390.

\bibitem[Galperin {\em et~al.\/}(2007)Galperin, Sukoriansky \&
  Anderson]{Galperin2007}
{\sc Galperin, B., Sukoriansky, S. \& Anderson, P.~S.} 2007 On the critical
  richardson number in stably stratified turbulence. {\em Atmospheric Science
  Letters\/} {\bf 8}~(3), 65--69.

\bibitem[Garrett \& Kunze(2007)]{Garrett2007}
{\sc Garrett, C. \& Kunze, E.} 2007 {Internal Tide Generation in the Deep
  Ocean}. {\em Annual Review of Fluid Mechanics\/} {\bf 39}~(1), 57--87.

\bibitem[Goff \& Arbic(2010)]{Goff2010}
{\sc Goff, J.~A. \& Arbic, B.~K.} 2010 {Global prediction of abyssal hill
  roughness statistics for use in ocean models from digital maps of
  paleo-spreading rate, paleo-ridge orientation, and sediment thickness}. {\em
  Ocean Modelling\/} {\bf 32}, 36--43.

\bibitem[Goff \& Jordan(1988)]{Goff1988}
{\sc Goff, J.~A. \& Jordan, T.~H.} 1988 {Stochastic Modeling of Seafloor
  Morphology: Inversion of Sea Beam Data for Second-Order Statistics}. {\em
  Journal of Geophysical Research\/} {\bf 93}~(B11), 13589.

\bibitem[Guo \& Holmes-Cerfon(2016)]{Guo2016}
{\sc Guo, Y. \& Holmes-Cerfon, M.} 2016 {Internal wave attractors over random,
  small-amplitude topography}. {\em Journal of Fluid Mechanics\/} {\bf 787},
  148--174.

\bibitem[Khatiwala(2003)]{Khatiwala2003}
{\sc Khatiwala, S.} 2003 {Generation of internal tides in an ocean of finite
  depth : analytical and numerical calculations}. {\em Deep Sea Research I\/}
  {\bf 50}, 3--21.

\bibitem[Ledwell {\em et~al.\/}(2000)Ledwell, Montgomery, Polzin, {St.
  Laurent}, Schmitt \& Toole]{Ledwell2000}
{\sc Ledwell, J.~R., Montgomery, E.~T., Polzin, K.~L., {St. Laurent}, L.~C.,
  Schmitt, R.~W. \& Toole, J.~M.} 2000 {Evidence for enhanced mixing over rough
  topography in the abyssal ocean}. {\em Nature\/} {\bf 403}, 179--182.

\bibitem[Lefauve {\em et~al.\/}(2015)Lefauve, Muller \& Melet]{Lefauve2015}
{\sc Lefauve, A., Muller, C. \& Melet, A.} 2015 {A three-dimensional map of
  tidal dissipation over abyssal hills}. {\em Journal of Geophysical Research:
  Oceans\/} {\bf 120}~(7), 4760--4777.

\bibitem[Legg(2014)]{Legg2014}
{\sc Legg, S.} 2014 {Scattering of Low-Mode Internal Waves at Finite Isolated
  Topography}. {\em Journal of Physical Oceanography\/} {\bf 44}, 359--383.

\bibitem[Li \& Mei(2014)]{Li2014}
{\sc Li, Y. \& Mei, C.~C.} 2014 {Scattering of internal tides by irregular
  bathymetry of large extent}. {\em Journal of Fluid Mechanics\/} {\bf 747},
  481--505.

\bibitem[Liu \& Yue(1998)]{Liu1998}
{\sc Liu, Y. \& Yue, D. K.~P.} 1998 {On generalized Bragg scattering of surface
  waves by bottom ripples}. {\em J. Fluid Mech.\/} {\bf 356}, 297--326.

\bibitem[{Llewellyn Smith} \& Young(2002)]{Llewellyn2002}
{\sc {Llewellyn Smith}, S.~G. \& Young, W.~R.} 2002 {Conversion of the
  Barotropic Tide}. {\em Journal of Physical Oceanography\/} {\bf 32}~(5),
  1554--1566.

\bibitem[Mack \& Schoeberlein(2004)]{Mack2004}
{\sc Mack, S.~A. \& Schoeberlein, H.~C.} 2004 Richardson number and ocean
  mixing: Towed chain observations. {\em Journal of Physical Oceanography\/}
  {\bf 34}~(4), 736--754.

\bibitem[Mathur {\em et~al.\/}(2014)Mathur, Carter \& Peacock]{Mathur2014}
{\sc Mathur, M., Carter, G.~S. \& Peacock, T.} 2014 Topographic scattering of
  the low-mode internal tide in the deep ocean. {\em Journal of Geophysical
  Research: Oceans\/} {\bf 119}~(4), 2165--2182.

\bibitem[Mei(1985)]{Mei1985}
{\sc Mei, C.~C.} 1985 {Resonant reflection of surface water waves by periodic
  sandbars}. {\em J. Fluid Mech.\/} {\bf 152}, 315--335.

\bibitem[M{\"{u}}ller \& Xu(1992)]{Muller1992}
{\sc M{\"{u}}ller, P. \& Xu, N.} 1992 {Scattering of Oceanic Internal Gravity
  Waves off Random Bottom Topography}. {\em Journal of Physical Oceanography\/}
  {\bf 22}~(5), 474--488.

\bibitem[Munk \& Wunsch(1998)]{Munk1998}
{\sc Munk, W. \& Wunsch, C.} 1998 {Abyssal recipes II: energetics of tidal and
  wind mixing}. {\em Deep-Sea Research: Part I\/} {\bf I}~(45), 1977--2010.

\bibitem[Nash {\em et~al.\/}(2007)Nash, Alford, Kunze, Martini \&
  Kelly]{Nash2007}
{\sc Nash, J.~D., Alford, M.~H., Kunze, E., Martini, K. \& Kelly, S.} 2007
  {Hotspots of deep ocean mixing on the Oregon continental slope}. {\em
  Geophysical Research Letters\/} {\bf 34}~(L01605), 1--6.

\bibitem[Nikurashin \& Ferrari(2010)]{Nikurashin2010a}
{\sc Nikurashin, M. \& Ferrari, R.} 2010 {Radiation and Dissipation of Internal
  Waves Generated by Geostrophic Motions Impinging on Small-Scale Topography:
  Application to the Southern Ocean}. {\em Journal of Physical Oceanography\/}
  {\bf 40}~(9), 2025--2042.

\bibitem[Polzin {\em et~al.\/}(1997)Polzin, Toole, Ledwell \&
  Schmitt]{Polzin1997}
{\sc Polzin, K.~L., Toole, J.~M., Ledwell, J.~R. \& Schmitt, R.~W.} 1997
  Spatial variability of turbulent mixing in the abyssal ocean. {\em Science\/}
  {\bf 276}~(5309), 93--96.

\bibitem[Sarkar \& Scotti(2017)]{Sarkar2017}
{\sc Sarkar, S. \& Scotti, A.} 2017 {From Topographic Internal Gravity Waves to
  Turbulence}. {\em Annual Review of Fluid Mechanics\/} {\bf 49}, 195--220.

\bibitem[Smith \& Sandwell(1997)]{Smith1997}
{\sc Smith, W.~H. \& Sandwell, D.~T.} 1997 {Global Sea Floor Topography from
  Satellite Altimetry and Ship Depth Soundings}. {\em Science\/} {\bf
  277}~(5334), 1956--1962.

\bibitem[{St Laurent} \& Garrett(2002)]{StLaurent2002}
{\sc {St Laurent}, L. \& Garrett, C.} 2002 {The Role of Internal Tides in
  Mixing the Deep Ocean}. {\em Journal of Physical Oceanography\/} {\bf 32},
  2882--2899.

\bibitem[Staquet \& Sommeria(2002)]{Staquet2002}
{\sc Staquet, C. \& Sommeria, J.} 2002 {Internal gravity waves : from
  instabilities to turbulence}. {\em Annu. Rev. Fluid Mech.\/} {\bf 34},
  559--593.

\bibitem[Sutherland(2010)]{Sutherland2010}
{\sc Sutherland, B.~R.} 2010 {\em {Internal Gravity Waves}\/}. Cambridge
  University Press.

\bibitem[Thurnherr {\em et~al.\/}(2005)Thurnherr, St~Laurent, Speer, Toole \&
  Ledwell]{Thurnherr2005}
{\sc Thurnherr, A.~M., St~Laurent, L.~C., Speer, K.~G., Toole, J.~M. \&
  Ledwell, J.~R.} 2005 {Mixing associated with sills in a canyon on the
  midocean ridge flank}. {\em Journal of Physical Oceanography\/} {\bf 35},
  1370--1381.

\bibitem[Timko {\em et~al.\/}(2017)Timko, Arbic, Goff, Ansong, Smith, Melet \&
  Wallcraft]{Timko2017}
{\sc Timko, P.~G., Arbic, B.~K., Goff, J.~A., Ansong, J.~K., Smith, W. H.~F.,
  Melet, A. \& Wallcraft, A.~J.} 2017 {Impact of synthetic abyssal hill
  roughness on resolved motions in numerical global ocean tide models}. {\em
  Ocean Modelling\/} {\bf 112}, 1--16.

\bibitem[Tobisch(2016)]{Tobisch2016}
{\sc Tobisch, E.} 2016 {\em {New Approaches to Nonlinear Waves}\/}, lecture no
  edn. Springer Heidelberg.

\bibitem[Waterhouse {\em et~al.\/}(2014)Waterhouse, MacKinnon, Nash, Alford,
  Kunze, Simmons, Polzin, {St. Laurent}, Sun, Pinkel, Talley, Whalen, Huussen,
  Carter, Fer, Waterman, Garabato, Sanford \& Lee]{Waterhouse2014}
{\sc Waterhouse, A.~F., MacKinnon, J.~A., Nash, J.~D., Alford, M.~H., Kunze,
  E., Simmons, H.~L., Polzin, K.~L., {St. Laurent}, L.~C., Sun, O.~M., Pinkel,
  R., Talley, L.~D., Whalen, C.~B., Huussen, T.~N., Carter, G.~S., Fer, I.,
  Waterman, S., Garabato, A. C.~N., Sanford, T.~B. \& Lee, C.~M.} 2014 {Global
  Patterns of Diapycnal Mixing from Measurements of the Turbulent Dissipation
  Rate}. {\em Journal of Physical Oceanography\/} {\bf 44}~(7), 1854--1872.

\bibitem[Wunsch \& Ferrari(2004)]{Wunsch2004}
{\sc Wunsch, C. \& Ferrari, R.} 2004 {Vertical Mixing, Energy, and the General
  Circulation of the Oceans}. {\em Annual Review of Fluid Mechanics\/} {\bf
  36}~(1), 281--314.

\bibitem[Yu \& Howard(2012)]{Yu2012a}
{\sc Yu, J. \& Howard, L.~N.} 2012 {Exact Floquet theory for waves over
  arbitrary periodic topographies}. {\em Journal of Fluid Mechanics\/} pp.
  1--20.

\bibitem[Yu \& Mei(2000{\natexlab{{\em a\/}}})]{Yu2000a}
{\sc Yu, J. \& Mei, C.~C.} 2000{\natexlab{{\em a\/}}} {Do longshore bars
  shelter the shore?} {\em Journal of Fluid Mechanics\/} {\bf 404}, 251--268.

\bibitem[Yu \& Mei(2000{\natexlab{{\em b\/}}})]{Yu2000b}
{\sc Yu, J. \& Mei, C.~C.} 2000{\natexlab{{\em b\/}}} {Formation of sand bars
  under surface waves}. {\em Journal of Fluid Mechanics\/} {\bf 416}, 315--348.

\bibitem[Zhao {\em et~al.\/}(2016)Zhao, Alford, Girton, Rainville \&
  Simmons]{Zhao2016}
{\sc Zhao, Z., Alford, M.~H., Girton, J.~B., Rainville, L. \& Simmons, H.~L.}
  2016 {Global Observations of Open-Ocean Mode-1 M2 Internal Tides}. {\em
  Journal of Physical Oceanography\/} {\bf 46}~(6), 1657--1684.

\bibitem[Zhao {\em et~al.\/}(2010)Zhao, Alford, MacKinnon \& Pinkel]{Zhao2010}
{\sc Zhao, Z., Alford, M.~H., MacKinnon, J.~A. \& Pinkel, R.} 2010 {Long-Range
  Propagation of the Semidiurnal Internal Tide from the Hawaiian Ridge}. {\em
  Journal of Physical Oceanography\/} {\bf 40}, 713--736.

\end{thebibliography}
%\bibliography{InternalChain}

\end{document}